\documentclass[useAMS,usenatbib]{mn2e}

\newcommand{\degree}{\hbox{$^\circ$}}
\newcommand{\ang}{\rm \r{A}}
\usepackage{graphics}
\usepackage{graphicx}
\usepackage{epstopdf}

\title[]{Stars and Dark Matter in the Spiral Gravitational Lens 2237+0305}

\author[C. M. Trott, T. Treu, L. V. E. Koopmans and R. L. Webster]{C. M. Trott$^{1}$\thanks{Email: ctrott@pet.mgh.harvard.edu}, T. Treu$^{2,4}$, L. V. E. Koopmans$^{3}$ and R. L. Webster$^{1}$\\
$^{1}$School of Physics, University of Melbourne, Victoria 3010,
Australia\\ $^{2}$Department of Physics, University of California,
Santa Barbara, CA 93106, USA\\ $^{3}$Kapteyn Astronomical Institute,
PO Box 800, 9700 AV Groningen, The Netherlands\\ $^{4}$Sloan Fellow,
Packard Fellow.}

\begin{document}

\maketitle

\begin{abstract}

We construct a mass model for the spiral lens galaxy 2237+0305, at
redshift $z_l$=0.04, based on gravitational-lensing constraints, HI
rotation, and new stellar-kinematic information, based on data taken
with the ESI spectrograph on the 10m Keck-II Telescope.  High
resolution rotation curves and velocity dispersion profiles along two
perpendicular directions, close to the major and minor axes of the
lens galaxy, were obtained by fitting the Mgb-Fe absorption line
region. The stellar rotation curve rises slowly and flattens at
$r\sim{1.5}\arcsec$ ($\sim$1.1 kpc). The velocity dispersion profile
is approximately flat. A combination of photometric, kinematic and
lensing information is used to construct a mass model for the four
major mass components of the system --- the dark matter halo, disc,
bulge, and bar. The best-fitting solution has a dark matter halo with
a logarithmic inner density slope of $\gamma$=0.9$\pm$0.3 for
$\rho_{\rm DM} \propto r^{-\gamma}$, a bulge with
M/L$_B$=6.6$\pm$0.3$\Upsilon_\odot$, and a disc with M/L$_B$
=1.2$\pm$0.3$\Upsilon_\odot$, in agreement with measurements
of late-type spirals.  The bulge dominates support in the inner
regions where the multiple images are located and is therefore tightly
constrained by the observations.  The disc is sub-maximal and
contributes $45\pm11$ per cent of the rotational support of the galaxy
at 2.2r$_d$. The halo mass is $(2.0\pm0.6)\times10^{12} M_{\odot}$, and
the stellar to virial mass ratio is $7.0\pm2.3$ per cent, consistent
with typical galaxies of the same mass.
\end{abstract}


\section{Introduction}

Strong gravitational lensing has been used extensively to model
galactic- and cluster-mass systems
\citep[e.g.][]{maller00,trott02,winn03,sand04,bradac06,halkola06,yoo06,jiang07,limousin07,bolton08b,grillo08}.
The combination of lensing information with photometric and kinematic
data can be used to break degeneracies inherent to each source of
information (e.g. the mass-sheet degeneracy and the mass-profile
anisotropy degeneracy) to produce a well-constrained mass model for
the system
\citep[e.g.][]{koopmans98,winn03,koopmans03,treu04,czoske08,koopmans06}. 
This is particularly important in spiral galaxies which have multiple
mass components that have to be modelled simultaneously, and the
limitations of individual methods are particularly
stringent. Specifically, in the absence of kinematic information, lens
image positions can accurately constrain the enclosed mass, and to
some extent also constrain the mass distribution up to a constant
mass-sheet \citep{gorenstein88,wambsganss94,wucknitz02}.  When assuming a stellar
mass-to-light ratio for the different optical components (e.g. disc
and bulge), based, for example, on stellar-population synthesis models, the dark matter contribution to the galaxy can then be inferred. In
the absence of lensing information, but with kinematic data in hand,
the overall galactic mass distribution is known, but again a stellar
mass-to-light ratio must be assumed for the optical components in
order to infer the contribution of the dark matter \citep[the
disc-halo degeneracy,][]{perez04,verheijen07}. Because of the rising
contribution of the stellar mass components, this becomes increasingly
more difficult towards the inner regions of the galaxy. With the
combination of lensing and high-resolution kinematic information,
however, the stellar mass-to-light ratios of the optical components
can be further constrained, breaking some of the degeneracies between
mass, mass-slope and anisotropy \citep[e.g.][and references
therein]{trott02,treu04}.

An accurate model for the mass distribution in spiral galaxies can
provide interesting information about both the baryonic and the
non-baryonic components. Disentangling these components provides
information about the mass-to-light ratios as well as the influence,
shape and density profile of the dark matter.  The main motivations of
this work are to: (1) break the classical degeneracy between the
stellar component and the dark-matter halo in spiral galaxies
\citep{vanalbada86}, using the additional information that lensing
provides; (2) determine independent mass-to-light ratios for the bulge
and disc components; (3) determine whether the disc component is
``maximal'' \cite[e.g.][]{sackett97,courteau99}. In more
general terms, understanding the relative mass of the different
components can help constrain models for disc galaxy formation
\citep[e.g.][and references therein, see also
\citet{shin07}]{dutton05,dutton07}. In addition, the presence of
massive discs in lens galaxies gives rise to interesting perturbations
to the standard elliptical models adopted for early-type galaxies with
a number of interesting consequences for the topology of the time
delay surface \citep{moller03}.

In this paper, we address these issues by studying the gravitational
lens galaxy 2237+0305.  Newly measured rotation curves and
line-of-sight velocity dispersion profiles along two directions close
to the major and minor axes -- obtained from Keck-ESI spectroscopy --
are combined with photometric and lensing data from HST to construct a detailed
mass model for the system. This is used to separate luminous and dark
matter over scales ranging from the central parts dominated by the
bulge out to several exponential scale lengths of the disc component,
with the aid of HI data.

The lens system 2237+0305 -- chosen for this study due to the low
redshift of the lens galaxy $z_l=0.04$ -- has been observed
extensively since its discovery in 1985 \citep{huchra85}. In addition
to providing information about the global mass structure of the galaxy
\citep{schneider88,kent88,mihov01,trott02}, it has also been used to
study the properties of stars, or compact objects, in the lensing
galaxy through microlensing \citep[e.g.][]{gilmerino05}, and the
structure of the broad-line region of the lensed quasar
\citep{wayth05,vakulik07,eigenbrod08a}. Previously published spectra of
2237+0305 have either concentrated on studying the quasar spectrum, or
only presented a small wavelength range for the galaxy spectrum. The
discovery paper, \citet{huchra85}, presented a quasar spectrum taken
with the MMT and featured both the CIV and CIII] broad lines. In
addition, they identified the H+K break at $\sim$4100$\ang$ and Mgb
absorption (5381$\ang$). This spectrum was not of sufficient quality
to identify any further features, but was successful in its primary
aim: to confirm the system as a lens. \citet{foltz92} observed the
galaxy to determine a velocity dispersion for the bulge. They
identified Mgb and FeI absorption features, but also only presented a
spectrum over a small wavelength range
(5100--5550$\ang$). \citet{lewis98} concentrated their observations
around the quasar emission lines to investigate variations between the
four images, thereby inferring microlensing effects (line and
continuum regions are expected to behave differently under
microlensing conditions because the flux emanates from regions of
different source size). The system 2237+0305 has also been the subject
of several microlensing monitoring experiments to study the size of
the quasar emitting regions \citep{wozniak00,udalski06,eigenbrod08b}.
\citet{rauch02} identified absorption lines in the
spectrum from intervening (between lens and source) objects using the
HIRES spectrograph on Keck. These spectra display high-resolution
absorption line profiles of the intervening systems and as such do not
concentrate on the spectrum of the galaxy as a whole. 

\citet{vandenven08} have recently obtained GMOS IFU data of the CaII
near-infrared triplet from the central region of 2237+0305 to
determine the contribution of dark matter in the central regions of
the galaxy. They demonstrated that within the central projected
4$\arcsec$, dark matter can contribute no more than 20\% of dynamical
mass, using single stellar population models to constrain the stellar
mass contribution, and both lensing and inferred kinematics to
constrain the total dynamical mass.  Our results agree with those of
\citet{vandenven08} in the region of overlap. However, our methodology
and aims are significantly different, as our goal is to construct a
simply parametrized model to describe the data over a tenfold radial
extent.

In a previous analysis of the system, \citet{trott02} combined lensing
and photometric data to constrain the inner mass distribution of the
galaxy. In addition, neutral hydrogen rotation points at large radius
were used to study the influence of the outer dark matter halo. They
found that additional kinematic information was required in the image
region to break the remaining model degeneracies. Specifically, the
inner logarithmic slope of the dark matter halo, a contentious issue
\citep[see for example][]{hayashi04,deblok05}, could not be constrained 
without additional information about the inner regions of the galaxy.

In contrast, this paper presents mass-to-light ratios for the bulge
and disc components that do not require stellar population-synthesis
models to constrain the stellar M/L ratios, and are therefore
\textit{independent} measurements. The only assumption is that the
mass-to-light ratio is constant across the component. This also allows
us to quantitatively assess the contribution and density profile of
the inner dark-matter halo in this high-mass spiral galaxy,
complementing work done in dwarf and low surface brightness spiral
galaxies
\citep{deblok01,borriello01,deblok05,zackrisson06}. Furthermore,
whereas in \citet{trott02} the halo was constrained to have a softened
isothermal shape due to the absence of kinematic data, in this paper
we have allowed for a more flexible model.

The paper is organized as follows. Section \ref{kinematics} describes
spectroscopic observations of 2237+0305 with ESI on Keck, presents the
resulting optical spectrum and uses absorption features to derive both
an optical rotation curve and line-of-sight velocity dispersion
profile along perpendicular axes. Section \ref{galaxy_DF} introduces
the modelling: Section \ref{mass_models} describes the mass density
models used for each component; a two-integral axisymmetric model of
the galaxy that is used to generate the observed kinematics is
introduced in Section \ref{kine_models}; then Section \ref{algorithm}
describes the algorithm used to find the solution.  Sections
\ref{results} and \ref{discussion} present the results of the
modelling and some discussion, and Section \ref{conclusion}
concludes. We assume $\Omega_{\rm m}=0.3$, $\Omega_{\Lambda}=0.7$ and
H$_0=70$ kms$^{-1}$Mpc$^{-1}$ throughout this paper.

\section{Spectroscopy and stellar kinematics}
\label{kinematics}

In this Section, the Echelle Spectrograph and Imager (ESI) on the
Keck-II telescope is used to measure a high spectral and spatial
resolution spectrum of the galaxy 2237+0305 and the lensed quasar
images.  The spectra are used to extract a rotation curve and velocity
dispersion profile for the galaxy, based upon measurements of the Mgb
absorption line. The results are consistent with measurements
\citep{vandenven08} in the region of overlap but extend them
considerably in extent.

\begin{figure*}
 \includegraphics[width=0.45\hsize]{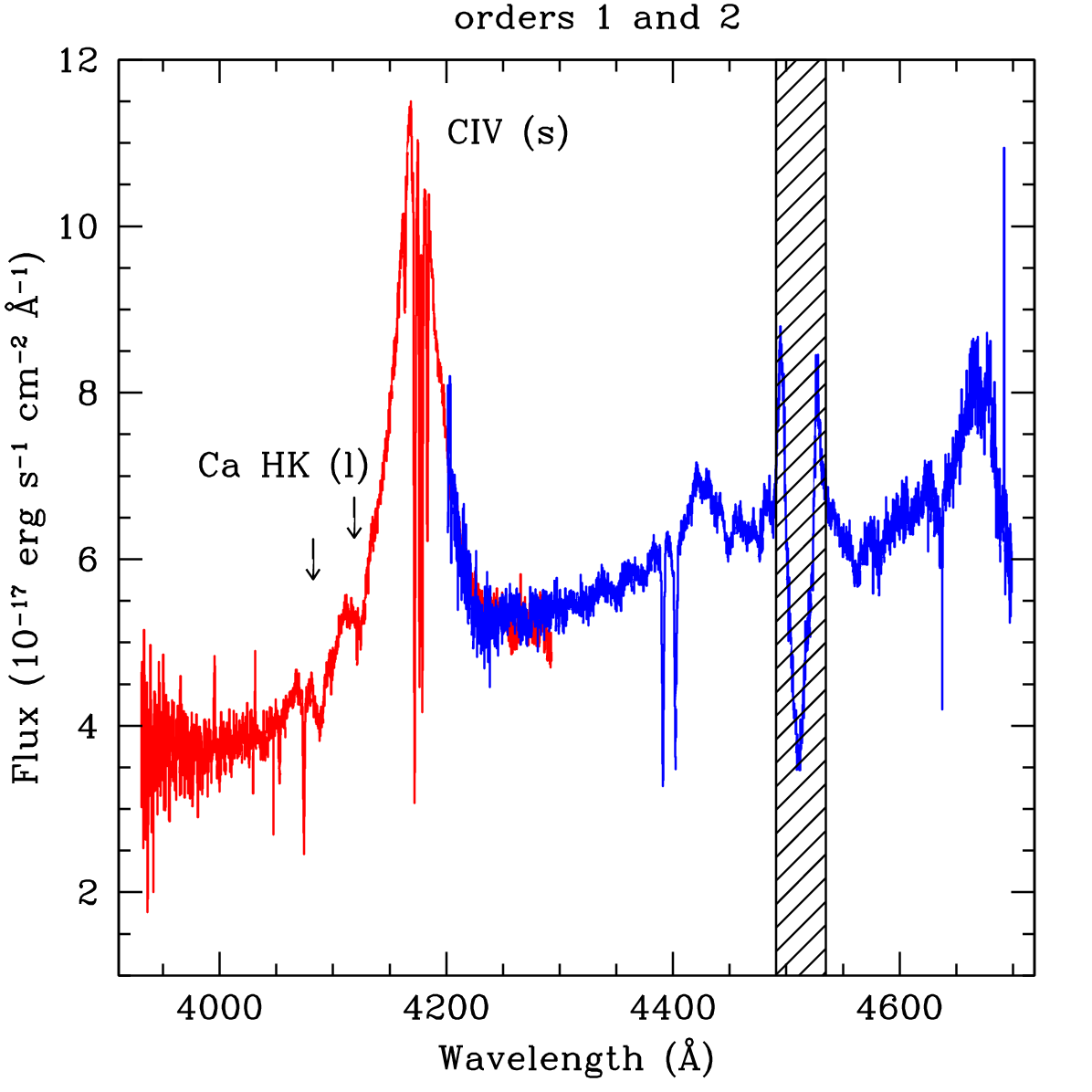}
 \hspace{0.3\fill}
 \includegraphics[width=0.45\hsize]{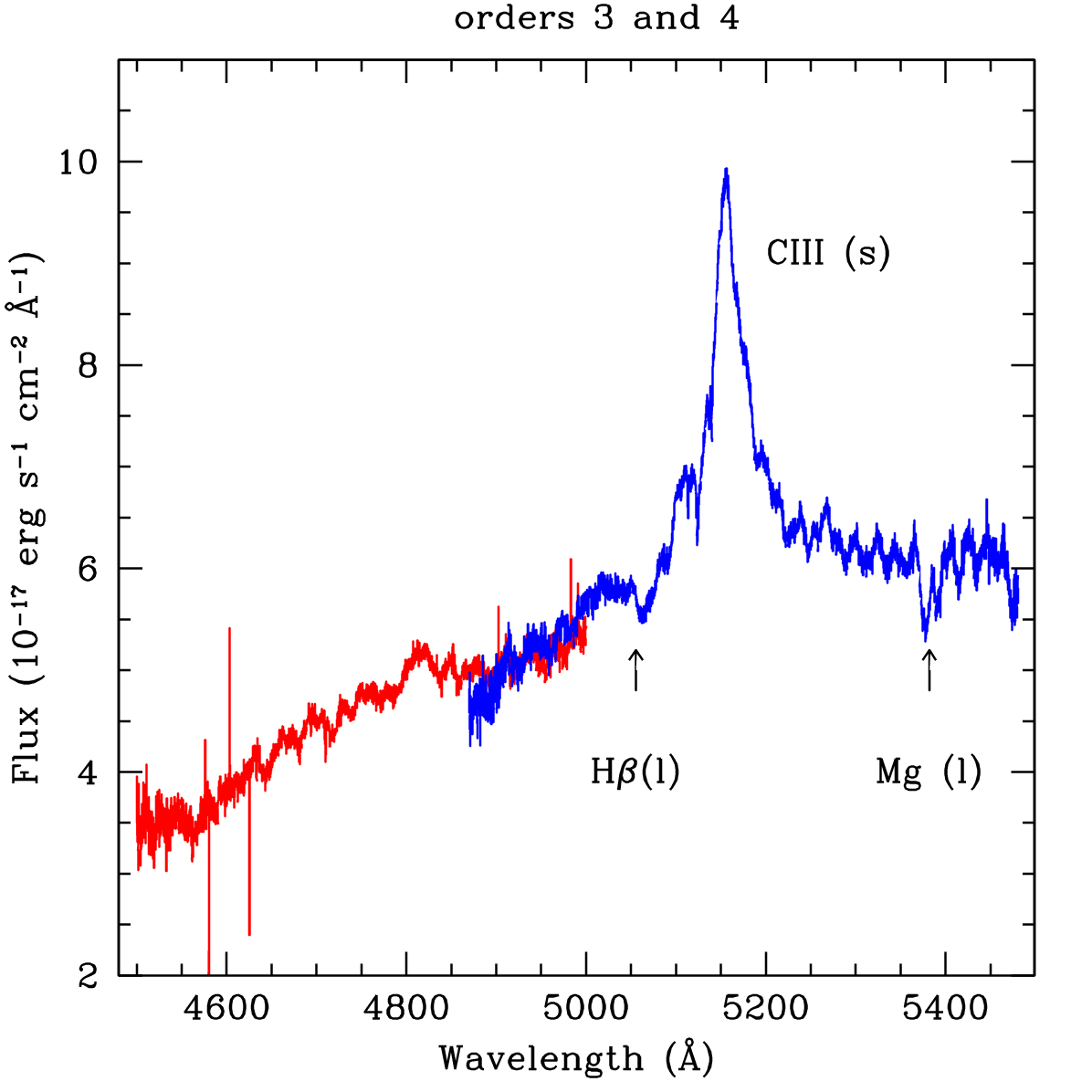}\\
 \vfil 
 \includegraphics[width=0.45\hsize]{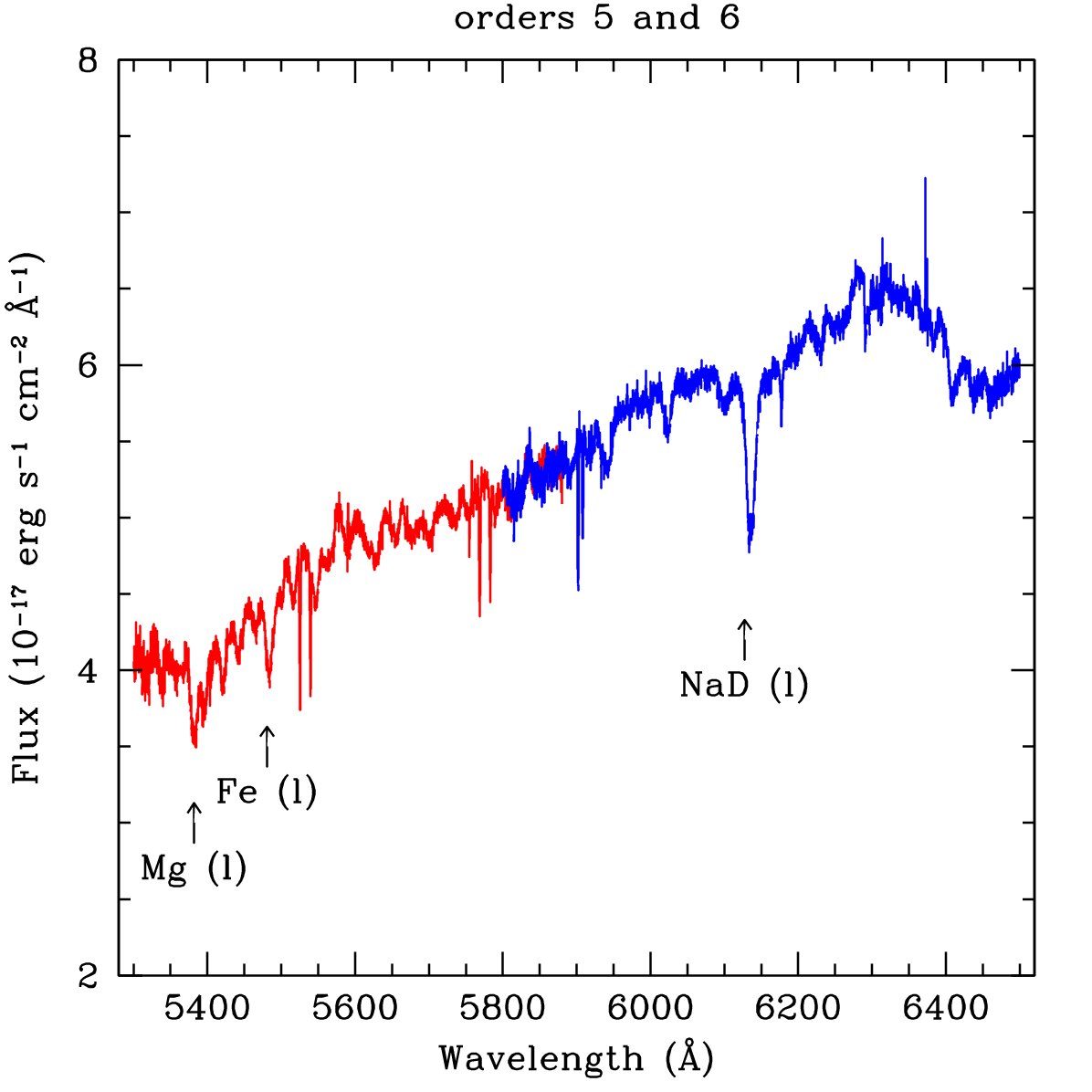}
 \hspace{0.3\fill}
 \includegraphics[width=0.45\hsize]{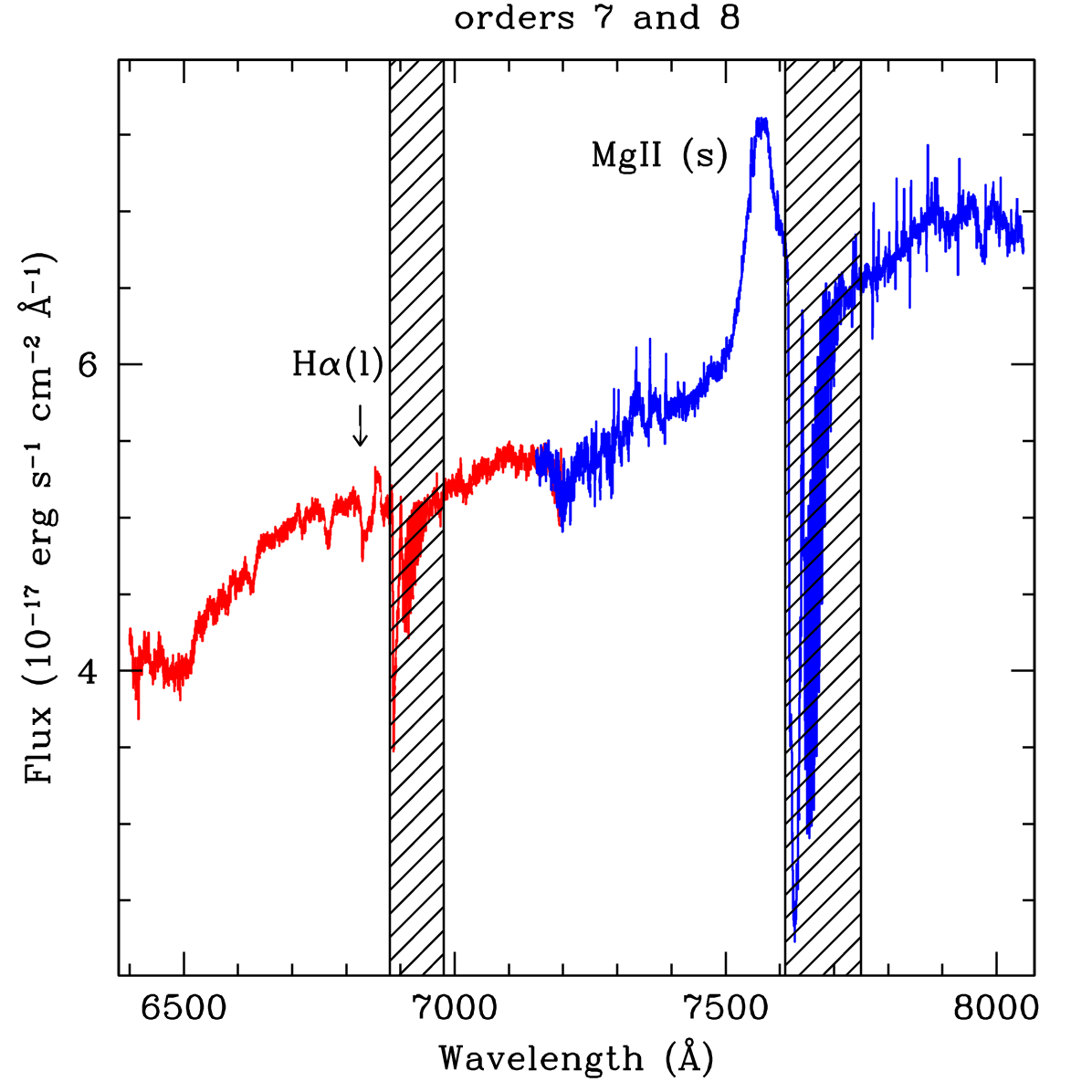}

\caption{Sky-subtracted spectra of 2237+0305 taken at a position angle
12$\degree$ clockwise from that of the major axis. The spectra have
been flux calibrated up to a constant factor based on ESI standard
response curve, measured in July 2001. Two orders are shown in each
plot (odd and even orders in red and blue respectively). Relevant
features are labelled for the lensed source (s) and for the lens
galaxy (l). A set of bad columns in order 2 and the atmospheric A and
B band in orders 7 and 8 are indicated by shaded
regions.}\label{order_all}
\end{figure*}

\subsection{Observations}\label{observations}

The lens galaxy 2237+0305 was observed as part of the Structure
and Dynamics Survey \citep[LSD, e.g. ][]{koopmans02}, a project that
aims to obtain stellar kinematic information in lens galaxies to
improve modelling and break degeneracies in their mass
models. The data were taken on 2001 July 21 and 23 with the Keck II
telescope on Mauna Kea, Hawaii. Conditions were clear and the seeing
was $0\farcs6$.  The Echelle Spectrograph and Imager (ESI) instrument
was used in echelle mode, providing a constant pixel size of 11.9
$\rm{kms^{-1}}$pixel$^{-1}$ across all orders, and a spatial
resolution along the slit of approximately 0.154 arcsec/pixel
(depending on the order). The galaxy was observed with a $1.25\arcsec$ slit width
 giving a kinematic resolution of $\Delta{v}$=36.1
$\rm{kms^{-1}}$.

The target was observed along two perpendicular slit directions,
sampling close to the lens galaxy major and minor axes and trying to
avoid the locations of the lensed quasar images. Table
\ref{observations_table} summarises the key observational data for the
galaxy. Note that the major axis is at a position angle of
77$\degree$ (measured as usual East from North). Although the
observations were taken at 12$\degree$ from the major and minor axes
to minimise quasar contamination, for simplicity the two slit angles
will be referred to as the `major' and `minor' axes herein. The
misalignments are taken into account in the modelling.  Two of the
spectra were observed with the galaxy offset 5$\arcsec$ from the slit
centre in order to sample the outer regions on one side.

\begin{table}
\centering
\small
\begin{tabular}{lcccc}\hline
{Date}&{PA}&{Slit Position}&{Slit Width}&{Exposure} \\\hline
{}&{$\degree$}&{$\arcsec$}&{$\arcsec$}&{Seconds} \\ \hline
{2001 July 21}&{65} &{5 (offset)}&{1.25}&{1200}\\
{2001 July 21}&{155}&{5 (offset)}&{1.25}&{1200}\\
{2001 July 21}&{155}&{centred}&{1.25}&{1200}\\
{2001 July 23}&{65} &{centred}&{1.25}&{1800}\\
\hline
\end{tabular}
\caption{Observing log. \label{observations_table}}
\end{table}
\normalsize

In addition to the galactic exposures, template stellar spectra are
required to measure the rotation and velocity dispersion of the
galactic spectral features. The bulge of the galaxy is primarily
observed in these observations, and so template spectra are needed of
old red stars expected in such an environment. The I-band images of
2237+0305 presented in \citet{yee88} show the central 12$\arcsec$ to
be dominated by old and red stars with ($g-r$)=0.58 \citep[V --
R=0.9,][]{windhorst91} consistent with Giant K0III--K5III stars.  A
set of appropriate stellar templates, described by \citet{sand04}, were observed during the same run and are used in this
analysis.

\subsection{Optical spectrum and lens galaxy kinematics}\label{spectrum}

Data reduction was performed using the {\tt EASI2D} software package,
developed by David J.~Sand and T.~Treu \citep{sand04}. Details of the
data reduction are given in Appendix \ref{data_reduction}.

Figure~\ref{order_all} displays the first eight of the ten spectral
orders (the remaining two orders do not show any relevant features of
the lens galaxy) obtained by averaging the central four arcseconds in
order to increase the signal-to-noise ratio. Four arcseconds
correspond approximately to the region over which S/N $>$ 5 per
unbinned spatial pixel. A relative flux calibration was
performed. Prominent spectral features are marked on the spectra. The
line-of-sight absorption systems reported in \citet{rauch02}, between
the lens galaxy and background quasar, are also confirmed in these
spectra ($z$=0.5656: MgI 4466$\ang$, MgII 4377$\ang$, FeII 4070$\ang$;
$z$=0.827: MgII 5108$\ang$, FeII 4867$\ang$). There are no obvious
galactic emission lines at the redshift of the lens, as expected from
an old stellar population. The H$\alpha$ absorption line (6821$\ang$
observed) is partly obscured by the B-band atmospheric water vapour
absorption line in lower resolution spectra, however our higher
resolution observations can separate the two lines and H$\alpha$
emission is not found to be present. The broad emission lines of the
lensed quasar images, CIV (4175$\ang$), CIII] (5145$\ang$) and MgII
(7541$\ang$) are prominent, with CIV and MgII suffering from sky absorption
\citep[see also ][]{huchra85}.

The spectra were used to
measure the rotation curve and line-of-sight velocity dispersion
profile for the major and minor axes as described in
Appendix~\ref{data_reduction}.  Figure \ref{best_fits} displays the
best fits and their uncertainties. The major-axis velocity dispersion
profile is relatively flat, immediately suggesting an underlying isothermal density
profile \citep[e.g.][]{bertin02}.

\begin{figure}
\includegraphics[width=8cm]{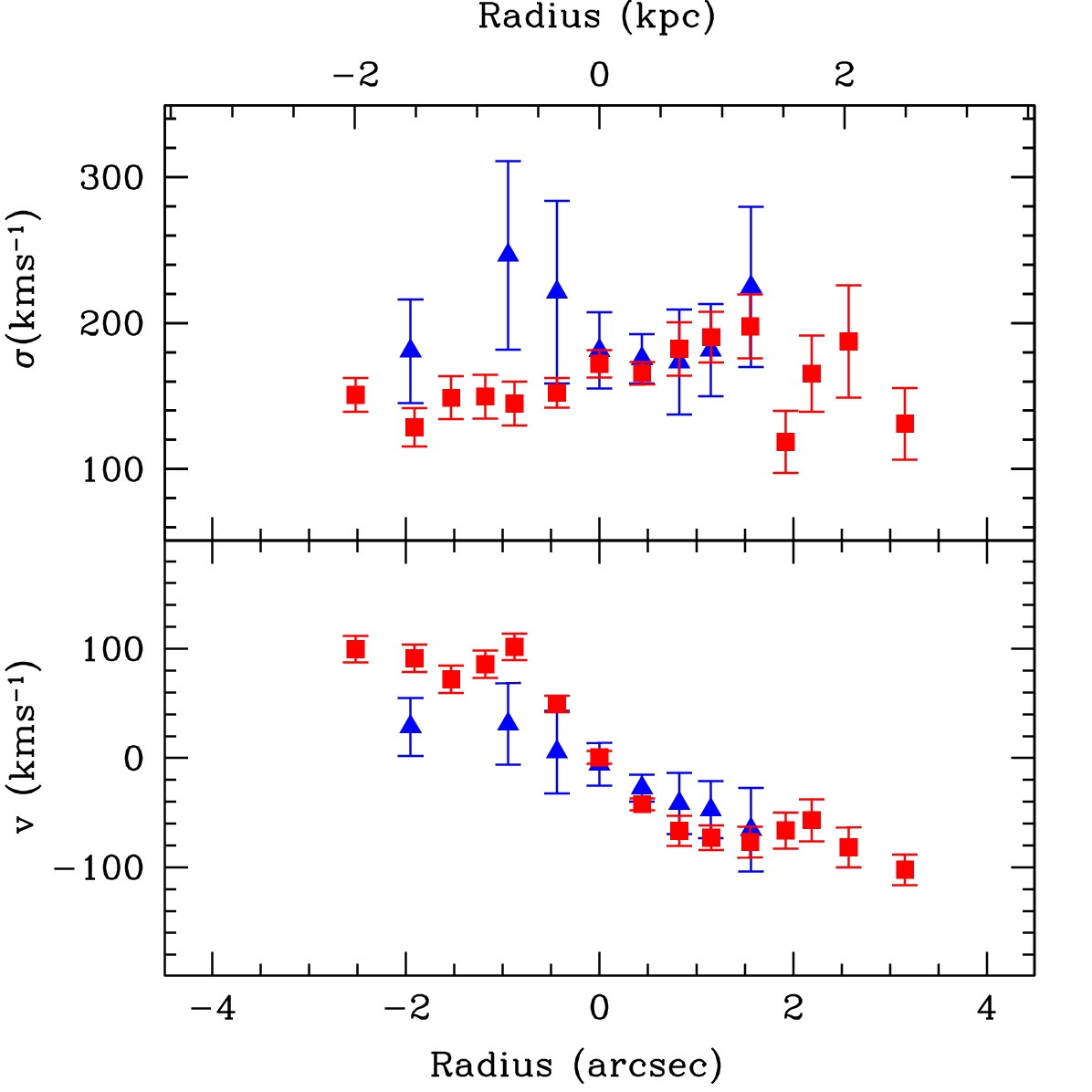}
\caption{Velocity dispersion profile and rotation curve for the major (red squares) and minor
(blue triangles) axes. The larger error
bars far from the galactic centre reflect the lower signal-to-noise in
these regions.}\label{best_fits}
\end{figure}

\section{The galaxy model}\label{galaxy_DF}

In \citet{trott02}, it was argued that kinematic information could
break the remaining degeneracies (see discussions above) between
models. No kinematic information for the galaxy centre was available
at that time, however, and degeneracies remained in the
modelling. With new high quality kinematic information in hand, we can
now further constrain the mass distribution.

\subsection{Mass models}\label{mass_models}

Given the larger number of constraints with the
addition of kinematic information, one can accommodate more
flexible models for the mass components than
\citet{trott02}. There is no reason to change the modelling of the
disc (exponential surface density) and bar (Ferrers ellipse ---
although the non-axisymmetric component of the bar will not be considered for the inner kinematic model),
since these are based on visible light profiles. For the kinematic
analysis, the exponential disc is deprojected to an exponential
term for the $z$-direction \citep{bintre87}. The distribution of dark
matter remains unknown, and a more general profile can be fitted than
was used in \citet{trott02}.

\subsubsection{Stellar component}

For tractability, we assume a model for the bulge component that is
simple and retains the key features
that are required both for the lensing and kinematic constraints. In previous work, and following that of \citet{schmidt96},
the bulge is modelled by a flattened de Vaucouleurs surface
brightness profile. The latter is one of a class of S\'{e}rsic
profiles \citep{sersic68}:
\begin{equation}
I(R) = I_0 \exp{[-b_m(R/r_b)^{1/m}]},
\end{equation}
$r_b$ is the characteristic scale length and $b_m$ is a
constant. A de Vaucouleurs profile corresponds to $m=4$, with an
appropriately scaled $r_b$.

\citet{limaneto99} and \citet{lokas04} present de-projections of the
circularly symmetric S\'{e}rsic profile, which allow one to use
kinematic expressions for the rotation speed and line-of-sight
velocity dispersion. We generalise these for a
non-spherical density profile, because a flattened bulge component is
important for the lensing.  The deprojected elliptical S\'{e}rsic profile for
general $m$ is given by,
\begin{eqnarray}
\rho(\xi) &=& \rho_0\left(\frac{\xi}{\alpha}\right)^{-p}\exp{\left[-\left(\frac{\xi}{\alpha}\right)^{1/m}\right]},\\
\rho_0 &=& \Sigma_0\frac{\Gamma(2m)}{2\alpha\Gamma[(3-p)m]},\\
p &=& 1.0 - 0.6097/m + 0.05463/m^2,\\
\alpha &=& r_b/(b_m)^m
\end{eqnarray}
where $\alpha$ is a generalised scale length, $\Sigma_0$ is the central
surface mass density ($\Sigma_0=\kappa_b\Sigma_{\rm cr}$ for connection with the
lensing, and $\Sigma_{\rm cr}$ is the critical surface mass density), and $\xi^2=r^2+z^2b^2/a^2$ is the elliptical radius, where $a$ and $b$ are the semi-major and -minor axes. We use the expressions of \citet{noordermeer08}
to describe the rotation curve of an elliptical S\'{e}rsic profile.

The scale length and ellipticity of the bulge were measured from a
two-dimensional fit of the HST NICMOS image (Proposal ID: 7495, PI:
Falco). After simulating the NICMOS point spread function with TINYTIM
\citep{hasan95}, a S\'{e}rsic plus point source (quasars) model was
fitted with the galfit software \citep{peng02}, yielding a S\'{e}rsic
index of $m$=4.0$\pm$0.1, a scale length of
$r_b$=4.07$\pm$0.10$\arcsec$ and a projected axis ratio of
$(a/b)_{b,pr}$=0.64$\pm$0.01. We use an intrinsic axis ratio of
$(a/b)_{b,in}$=0.6 \citep{vandenven08}.

\subsubsection{Dark matter component}

The dark matter halo is modelled with a spherical generalised cusp, with variable inner and outer density slope
\citep{zhao96,munoz01,keeton01}. This profile reduces to a variety of
useful models, such as the NFW, Hernquist and isothermal, and has the
form,
\begin{equation}
\rho(r) = \frac{\rho_s}{(r/r_h)^\gamma[1 + (r/r_h)^2]^{(n-\gamma)/2}},
\end{equation}
where $\gamma$ and $n$ are the inner and outer logarithmic slopes,
respectively, and $r_h$ is the characteristic scale length. For a
spherical model, the lensing convergence, $\kappa$, is given by,
\begin{eqnarray}
\kappa(x) = \kappa_h B\left(\frac{n-1}{2},\frac{1}{2}\right)(1 + x^2)^{(1-n)/2}  \times \nonumber \\
	{}_2F_1\left[\frac{n-1}{2},\frac{\gamma}{2},\frac{n}{2};\frac{1}{1+x^2}\right],
\end{eqnarray}
where $\kappa_h = \rho_sr_h/\Sigma_{\rm cr}$, $x = r/r_h$, $B(\dots)$
is the beta function and $_2F_1(\dots)$ is a hypergeometric
function \citep{bucholz69}. In our model, the outer logarithmic slope, $n$, will not
be well constrained as most of the constraints lie
within 5$\arcsec$ of the galaxy centre. It is set to a value of
3, consistent with N-body simulations
\citep{nfw96,moore99,nav04}.

\subsection{Kinematic modelling}\label{kine_models}

To model the lensing and kinematic data-sets of 2237+0305, we assume
that the central regions of the galaxy can be well-approximated by an
axisymmetric two-integral distribution.

We solve the two-integral Jean's equations from which we obtain the
velocity dispersion profile and mean streaming motion of the galaxy.
We follow the prescription of \citet{vandermarel07}, details of which
are described in Appendix~\ref{axisymmetric}, and solve the
two-integral axisymmetric Jean's equations on a grid. We model the
bulge+bar as one single axisymmetric component, ignoring the
second-order non-axisymmetric bar component.

Jean's equations provide the first two dynamical
moments,
\begin{equation}
\langle\overline{v_w^m}\rangle(x,y) = \frac{1}{\Sigma(x,y)}\int^{\infty}_{-\infty}\rho(x,y,w)\overline{v_w^m}(x,y,w)dw,
\end{equation}
where $w$ denotes the line-of-sight direction and
\begin{eqnarray}
\overline{v_w} &=& \overline{v_\phi}\cos{\phi}\sin{i},\\\nonumber
\overline{v_w^2} &=& (\overline{v_\phi^2} - \overline{v_R^2})\cos^2\phi\sin^2i + \overline{v_R^2}
\end{eqnarray}
are the first and second moments, $i$ is the inclination
angle, and $R$ and $\phi$ are the radial and azimuthal directions. The rotation component, $V$, and velocity dispersion, $\sigma$,
are then derived as,
\begin{eqnarray}
V &=& <\overline{v_w}>,\\\nonumber
\sigma^2 &=& V^2 - \sigma^2_{RMS}\\
&=& V^2 - <\overline{v_w^2}>.
\end{eqnarray}

The effects of seeing, pixel size and slit width are incorporated into
the modelling using the Monte-Carlo technique of
\citet{vandermarel94}. The luminosity-weighted moments are convolved
with the point spread functions that characterise the slit width,
pixel size and seeing. This is accomplished by randomly sampling the
region of the galaxy in question and averaging luminosity-weighted
samples: a point is chosen randomly (uniform sampling) from within the
slit and pixel, and then a normal deviate (with FWHM equal to the
seeing) is added. The kinematics at this point are computed and added
(with a weight proportional to the galaxy luminosity at that point) to
previous samples. This method produces good results without having to
perform expensive numerical convolutions. We sample 3000 points for
each datum to arrive at our observed moments. By computing
kinematics with a larger number of samples and comparing output, we
find that 3000 samples are sufficient to produce accurate results (error $<$10\% of measurement error).

\subsection{Mass-model optimization}\label{algorithm}

\begin{table}
\small
\centering
\begin{tabular}{c||c|c|c}
{Component} & {Parameter} & {Value} & {Definition}\\\hline\hline
{Bulge} & {$\kappa_b$} & {free} & {convergence}\\
{} & {$r_b$} & {4.07$\arcsec$} & {scale length}\\
{} & {$a/b_{b,pr}$} & {0.64} & {projected axis ratio}\\
{Rotation} & {$k$} & {free} & {rotational support}\\\hline
{Disc} & {$\kappa_d$} & {free} & {convergence}\\
{} & {$r_d$} & {11.3$\arcsec$} & {scale length}\\
{} & {$a/b_d$} & {0.5} & {axis ratio}\\\hline
{DM Halo} & {$\kappa_h$} & {free} & {convergence}\\
{} & {$r_h$} & {free} & {scale length}\\
{} & {$\gamma$} & {free} & {inner slope}\\
{} & {$n$} & {3.0} & {outer slope}\\\hline
{Bar} & {$\kappa_{br}$} & {free} & {convergence}\\\hline
{Source} & {$\beta_x$} & {free} & {$x$ source position}\\
{} & {$\beta_y$} & {free} & {$y$ source position}\\\hline
\end{tabular}
\caption{Parameters to be fitted in the combined lensing and kinematic model of 2237+0305. The ellipticities of the bulge and disc are fixed at their photometric value, and the bar and disc scale lengths are modelled as in \citet{schmidt96} with only a variable M/L. There are 9 parameters in total.}\label{params_new_2237}
\end{table}

The algorithm for consistently solving for the kinematics and lensing
constraints involves a simplex minimisation of the $\chi^2$ of the
mass model. Specifically, we: (1) construct a galaxy mass model as
described in Section \ref{mass_models} with parameters shown in Table
\ref{params_new_2237}; (2) calculate the lens image positions based on
this model; (3) calculate the observed rotation and velocity
dispersion profile based on a summation of the disc, bulge and dark
matter halo components; (4) calculate the expected rotation at the
location of the outer HI measurements from all mass components; (5)
compute a $\chi^2$ for the model as a summation of the adequacy of
fits for the lens positions, inner kinematics and outer kinematics;
(6) iteratively change the parameters until an acceptable minimum is
obtained.

The parameters in the model are given in Table~\ref{params_new_2237},
including the unknown source position. We use the HST-derived image
positions of \cite{crane91}, as in \cite{trott02}.

\normalsize The constraints\footnote{Flux ratios were not used to
constrain the models due to the added complexities of modelling
microlensing, substructure and the effects of differential dust
extinction, however we do calculate the resultant ratios and compare
them with mid-IR measurements as a model sanity check.} include the
four image positions, the two HI rotation points and the kinematic
information, with points in the minor axis rotation curve removed due
to contamination, giving 54 constraints, and 45 degrees of freedom.

The $\chi^2$ statistic quantifying the goodness-of-fit for a model is
defined as,
\begin{eqnarray}
\chi^2 &=& \sum_{\rm Images}\frac{(\theta-\theta_{\rm
 mod})^2}{\sigma_{\rm images}^2} + \sum_{\rm r_{HI}}\frac{(v-v_{\rm
 mod})^2}{\Delta{v}^2}\\\nonumber &+& \sum_{\rm
 r_{Keck}}\frac{(v-v_{\rm mod})^2}{\sigma_{v,\rm data}^2} + \sum_{\rm
 r_{Keck}}\frac{(\sigma_{\rm data}-\sigma_{\rm mod})^2}{\sigma_{\rm
 data}^2},
\end{eqnarray}
where the summations are over lens image positions, HI rotation
points, Keck major and minor axis rotational values and Keck major and
minor axis velocity dispersion values, respectively. The reduced
$\chi^2$, given the large number of degrees of freedom, is defined as
$\chi^2$/45. A reduced $\chi^2$ of unity is a statistically acceptable
fit. Image position errors are 8mas, consistent with measurement
errors.

\section{Results}\label{results}

\begin{table}
\small
\centering
\begin{tabular}{c||c|c}
{Component} & {Parameter} & {Value}\\\hline\hline
{Bulge} & {$M/L_B$} &  {6.6$\pm$0.3$\Upsilon_\odot$}\\
{} & {$\kappa_b$} & {103$\pm$5}\\
{} & {$r_b$} & {$\equiv$4.07$\arcsec$}\\
{} & {$k$} & {1.0$\pm$0.1}\\\hline
{Disc} & {$M/L_B$} & {1.2$\pm$0.3$\Upsilon_\odot$}\\
{} & {$\kappa_d$} & {0.014$\pm$0.004}\\
{} & {$r_d$} & {$\equiv$11.3$\arcsec$}\\\hline
{DM Halo} & {$\kappa_h$} & {0.010$\pm$0.003}\\
{} & {$r_h$} & {(31.7$^{+15.0}_{-9.0}$)$\arcsec$}\\
{} & {$\gamma$} & {0.9$\pm$0.3}\\
{} & {$n$} & {$\equiv$3.0}\\\hline
{Bar} & {$\kappa_{\rm br}$} & {0.06$\pm$0.01}\\\hline
{Source} & {$\beta_{\Delta{E}}$} & {-0.058$\pm$0.006$\arcsec$}\\
{} & {$\beta_{\Delta{N}}$} & {-0.015$\pm$0.006$\arcsec$}\\\hline
{Magnification} & {Image A} & {4.3}\\
{} & {Image B} & {4.4}\\
{} & {Image C} & {-2.4}\\
{} & {Image D} & {-5.1}\\
{} & {$\mu_{\rm tot}$} & {16.2}\\\hline\hline
{$\chi^2$} & {Reduced $\chi^2$} & {2.2}\\
\end{tabular}
\caption[Parameter values for the best-fitting solution with reduced $\chi^2$ = 2.2.]{Parameter values for the best-fitting solution with reduced $\chi^2$ = 2.2. Here, $\beta_{\Delta{E}}$ and $\beta_{\Delta{N}}$ refer to the source position relative to the galactic centre. Uncertainties are also shown corresponding to an increase of one in the $\chi^2$, while marginalising over the remaining parameters. \label{best_fit_rc_no_prior_table}}
\end{table}

\begin{figure}
\begin{center}
\includegraphics[width=0.9\hsize]{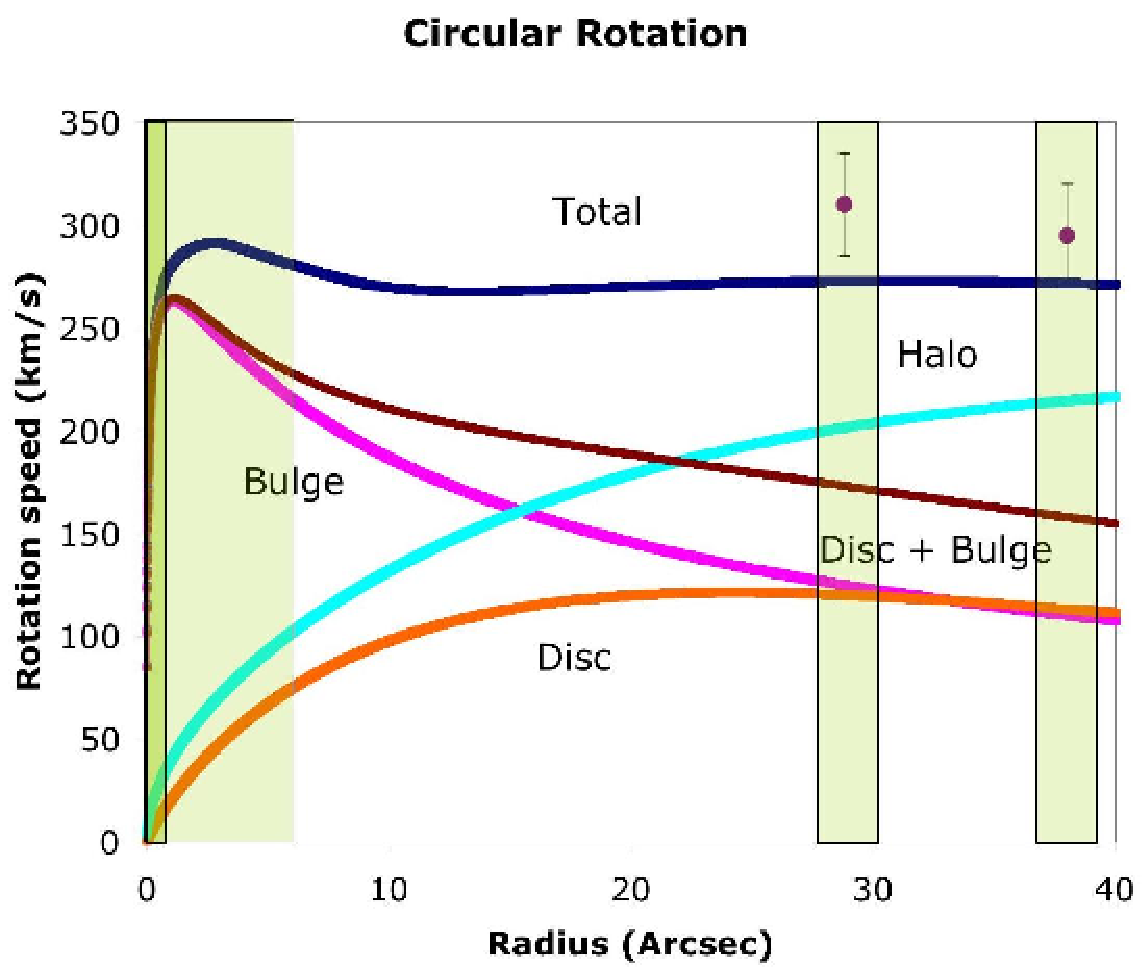}\\
\includegraphics[width=0.9\hsize]{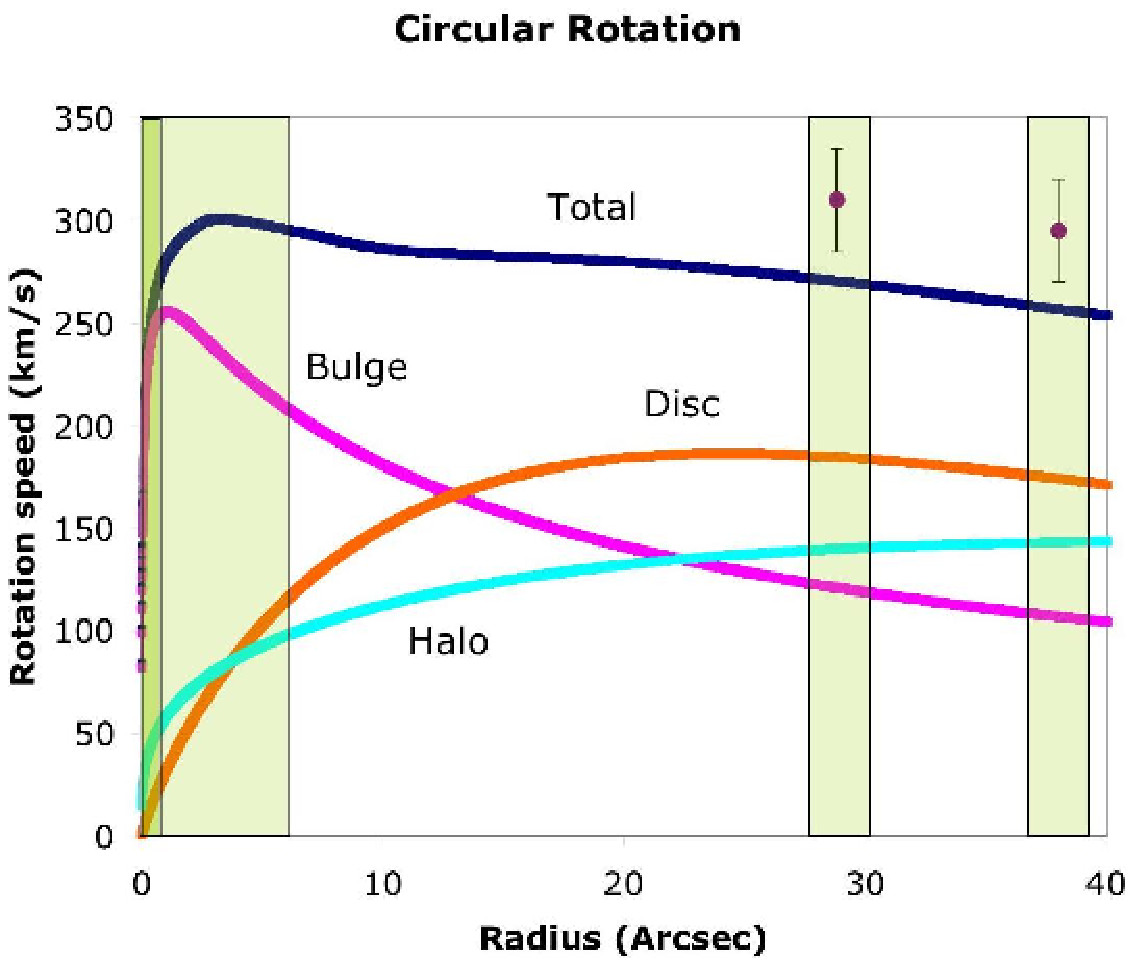}
\caption[Rotation curve for the solution with the best $\chi^2$.]{(Upper) Rotation curve (edge-on) for the solution with the best $\chi^2$. Light-shaded regions indicate radii where kinematic information is available. The dark-shaded region in the galactic centre indicates where the lensing information is useful. The two HI points are shown with their $1\sigma$ measurement uncertainties. The contribution of the luminous component alone (disc+bulge) is also displayed. (Lower) Rotation curve for a solution with minimum dark matter halo. This solution is allowed within the 3$\sigma$ error bars and demonstrates the disc-halo degeneracy in the model. \label{best_fit_rc_no_prior}}
\end{center}
\end{figure}

Table \ref{best_fit_rc_no_prior_table} lists the parameters for the
best-fitting model and the 1 $\sigma$ uncertainties for each parameter, defined
as the parameter value corresponding to an increase
in the $\chi^2$ of one while marginalising over the other
parameters. This is achieved by sampling select points in parameter space, and determining where the $\chi^2$ increased by one. Specifically, to determine the uncertainty on parameter, alpha, for example, we performed a full minimization for several values of alpha (i.e., set the value of alpha and allow all other parameters to vary to find the minimum) and find the value for alpha where the $\chi^2$ value increased by one from the global minimum.

Figure \ref{best_fit_rc_no_prior} (upper) displays the
circular velocity curve for the best model.
Also displayed are regions of the galaxy where kinematic or lensing
information adds constraints.  As expected, the model with the
best-fitting solution is bulge-dominated in the central regions and
has a stellar disc that, combined with the bulge, supports the
majority of the rotation to $r\sim$ 20$\arcsec$, beyond which the dark
matter halo dominates.  The inner dark matter halo slope is consistent
both with an NFW profile and a central core at 3$\sigma$. The bulge
and disc mass-to-light ratios are consistent with previous
measurements. Table \ref{best_fit_rc_no_prior_table} also displays the
image magnifications and total magnification for the model. The flux
ratios are consistent with the mid-IR measurements of \citet{agol00},
providing a good sanity check for the model.

\begin{figure}
\begin{center}
\includegraphics[width=\hsize]{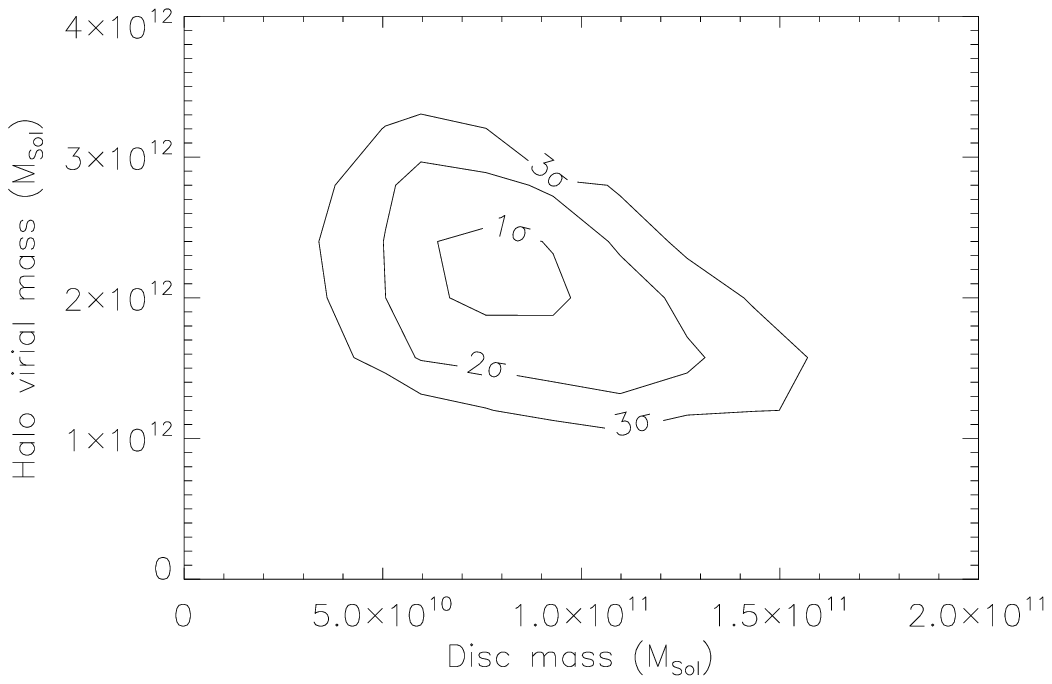}\\
\caption[]{Degeneracy between the disc mass and halo virial mass, with contours at 1, 2 and 3$\sigma$ above the minimum (minimum $\chi^2$+2.3, 6.2, 9.2). \label{disk_halo_degen}}
\end{center}
\end{figure}

Figure \ref{best_fit_rc_no_prior} (lower) displays the rotation curve
of a solution that is at the very edge of what is allowed by the
uncertainties (3$\sigma$), having a minimum dark matter halo and a
more prominent disc. The relatively large errors on the HI data
points, the freedom of the dark matter model and the lack of kinematic
information between 5 and 30 arcseconds, combine to allow at 1\% level
a low-mass halo model. This demonstrates that the familiar disc-halo
degeneracy, where the disc convergence and shear can account for the
halo contribution, is almost completely broken. Figure \ref{disk_halo_degen} demonstrates the degree of degeneracy between the disc mass and halo virial mass. As expected, these parameters are correlated, but the system does not allow a large range for either. In a system with a larger disc-halo degeneracy, the range of disc and halo kappa values would be increased.
In 2237+0305, the
bulge contributes the majority of the lensing convergence and shear,
with the disc and bar both also contributing convergence and shear,
and the halo contributing convergence alone (spherical
model). Although the bulge cannot change to account for the disc
shear, there is freedom in the bar model to account for this. As such,
2237+0305 is a good system to break the disc-halo degeneracy. Even
tighter constraints can be obtained in the future if kinematic
information is available at a larger range of radii (notice that the
shape of the total rotation curve is different between the two
figures).  Future studies, including a prior on the stellar
mass-to-light ratio of the bulge and improved kinematic data, should
be able to further limit the remaining degeneracy between the halo and
disc mass.
\begin{figure*}
 \includegraphics[width=0.45\hsize]{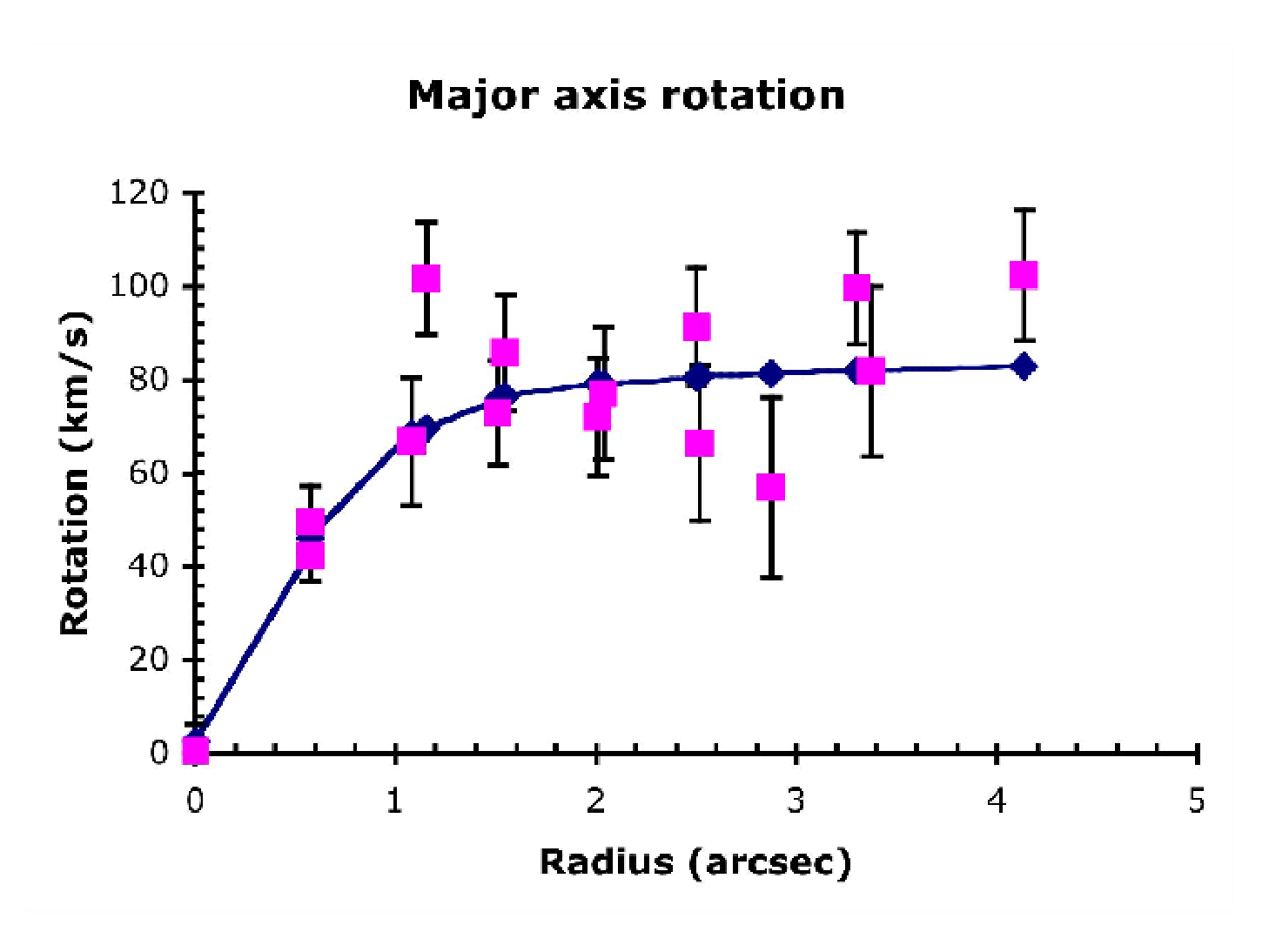}
 \hspace{0.3\fill}
 \includegraphics[width=0.45\hsize]{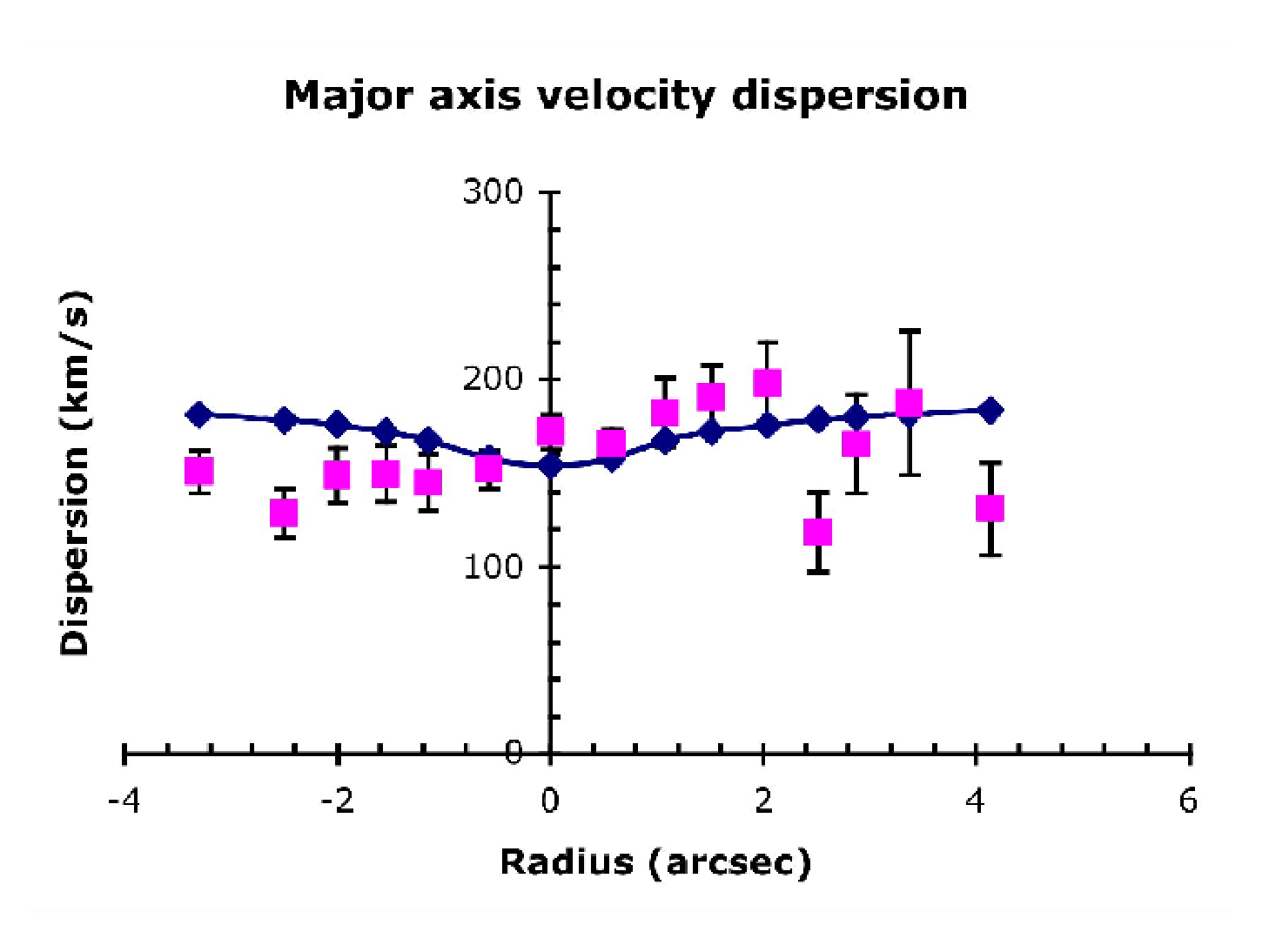}\\
 \vfil 
 \includegraphics[width=0.45\hsize]{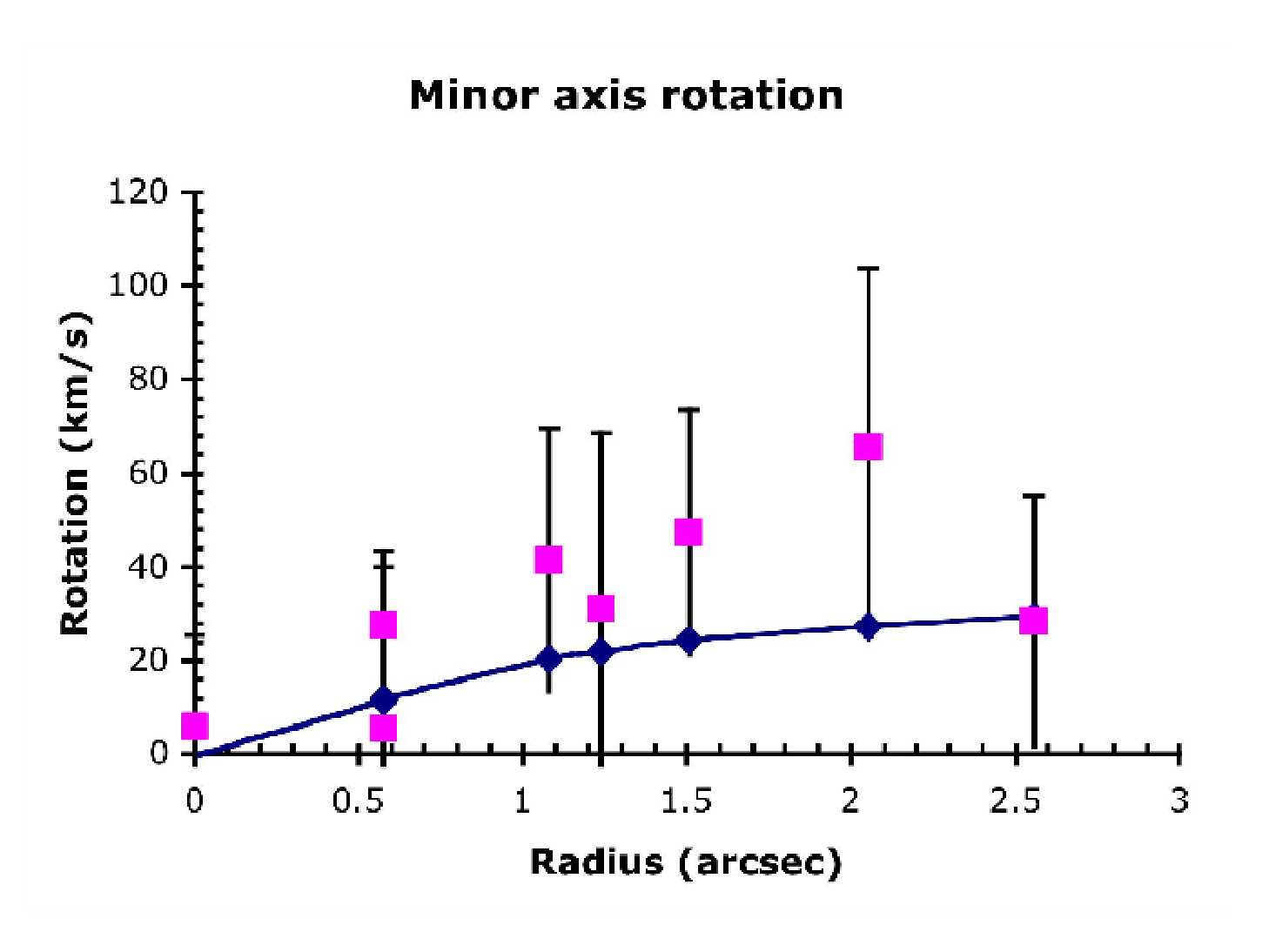}
 \hspace{0.3\fill}
 \includegraphics[width=0.45\hsize]{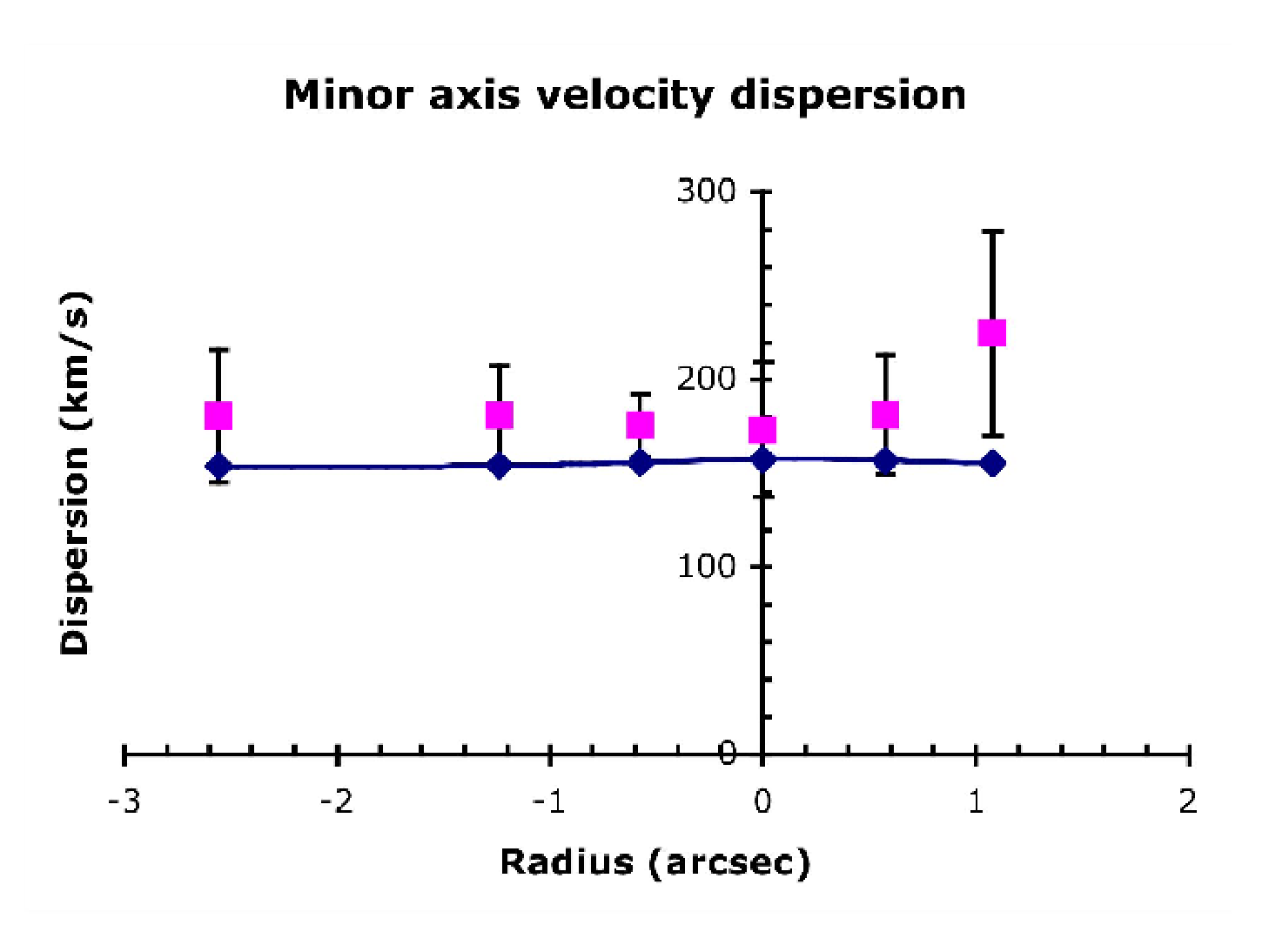}
\caption[Major axis data and fits for the rotation curve and
line-of-sight velocity dispersion profile for the best-fitting
solution.]{(Left) Major (upper) and minor (lower) axis data (pink
squares) and fits (blue diamonds) for the rotation curve (inclined and
calculated over an aperture) for the best-fitting solution. The
measurement uncertainties on the data are also shown as error
bars. (Right) Major (upper) and minor (lower) axis model and data
line-of-sight velocity dispersion profiles for the best-fitting
solution.}\label{best_fit_inner}
\end{figure*}

We emphasize, however, that there are correlations between other modelled parameters. The bulge, being the dominant component for the lensing, is well-constrained and displays only mild correlations with both the disc and halo kappa (results not shown), supporting our assertion that the disc-halo degeneracy is the most prominent. In terms of halo parameters, the halo convergence ($\kappa_h$) and inner slope ($\gamma$) display an anti-correlation, as do $\kappa_h$ and scale length ($r_h$). Although these correlations do exist, the small error bars on the bulge convergence tightly constrain the overall model. We note that there is freedom in the halo model, but this is to be expected given that there are no independent constraints on its shape. This freedom in the halo model is reflected in the relatively large error bars quoted on its parameters. By marginalizing over remaining parameters, the error bars include the effects of these correlations between parameters.

The model fits to the kinematics are also of interest. Figure
\ref{best_fit_inner} shows the fits to the inner regions.  The major
axis kinematics are well recovered, within the observational errors,
but the minor axis dispersion and rotation appear to be systematically
low. It is possible that the omission of the bar in the central
kinematics has contributed to this discrepancy.

\begin{figure}
\begin{center}
\includegraphics[width=\hsize]{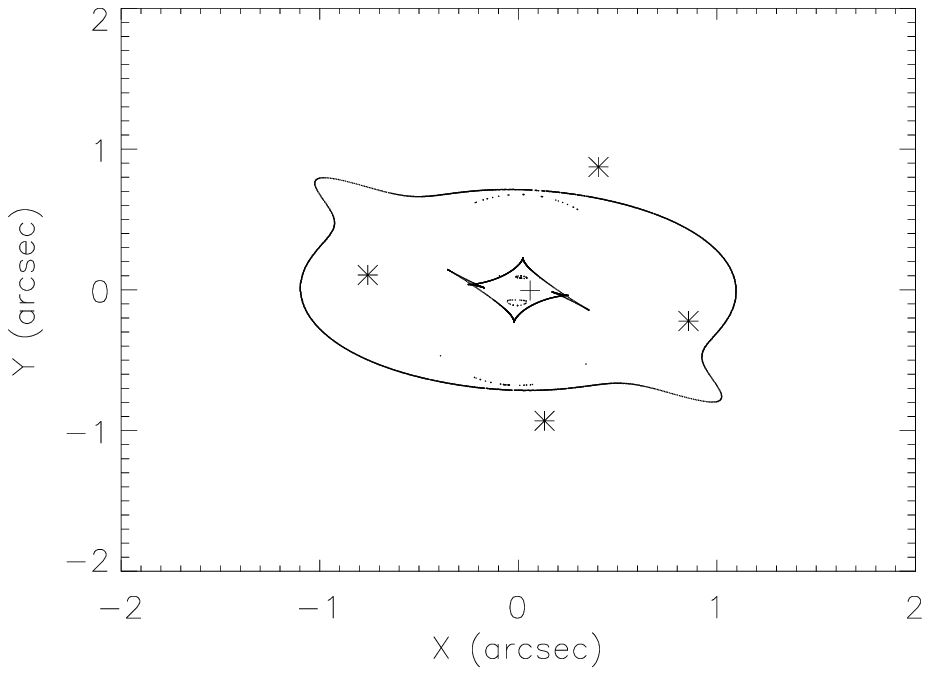}\\
\caption[]{Critical lines and caustic for the best-fitting model. Also shown are the image positions (asterisks) and source position (plus). Note that both radial and tangential caustics are displayed. \label{critical_curve}}
\end{center}
\end{figure}

Figure \ref{critical_curve} displays the critical lines and caustic for the best-fitting model.
The swallow-tail caustic feature and consequent non-elliptical critical line are due to the highly elliptical bar component. Note that the source does not lie close to the swallow-tail caustic and thus the details of the bar component do not strongly affect the image positions and magnifications.

Overall, the best-fitting model reproduces the data well considering
the coarseness of the data and models. The kinematics are reproduced
well with a kinematic best $\chi^2$ of 76 for 44 kinematic data points
(with 7 parameters to be fitted, giving 37 degrees of freedom). The
lensing reduced $\chi^2$ is somewhat higher than unity, as expected
because in general simple smooth models are not expected to reproduce
the image configuration with astrometric precision. However, this
level of precision is more than sufficient to determine robustly the
total enclosed mass within the Einstein radius
\citep{kochanek91,treu04}, the physical quantity that we are mostly
interested in for our model. The mass within the Einstein radius
(0.88$\arcsec$) is (1.67$\pm$0.07)$\times$10$^{10}$h$_{70}$M$_\odot$, consistent
with previous measurements \citep{wambsganss94,vandenven08}. The components of this enclosed mass are described in Table \ref{mass_enclosed}.
\begin{table}
\centering
\small
\begin{tabular}{lc}\hline
{Component}&{Mass enclosed}\\
{} & {$\times$10$^8$ M$_\odot$}\\\hline
{Halo} & {8.8$\pm$2.6}\\
{Bulge} & {142.5$\pm$7.0}\\
{Disc} & {4.1$\pm$1.0}\\
{Bar} & {11.6$\pm$1.9}\\
\hline
\end{tabular}
\caption{Mass enclosed within the Einstein radius (0.88$\arcsec$) for each of the four mass components. \label{mass_enclosed}}
\end{table}
\normalsize
The components of the $\chi^2$ are as follows: the lens image positions
contribute 21.0\% (representing 14.8\% of the constraints), the HI
points contribute 3.0\% (3.7\%), the inner rotation contributes 20.6\%
(42.6\%), and the inner dispersion contributes 55.4\% (38.9\%).

\subsection{Disc maximality and global parameters}

With the best fitting model in hand we can now try to address the
question of whether the stellar disc of the lens galaxy 2237+0305 is
maximal or not. The relevant quantity for comparison with other
observations and theory is V$_{2.2}/$V$_{\rm tot}$, i.e. the ratio
between the disc rotation velocity and the total rotation velocity
measured at 2.2 scale lengths, close to the peak of the disc rotation
curve.  Based on the best fitting model, and taking into account the
errors on the best fitting parameters, we find that V$_{2.2}/$V$_{\rm
tot}$ = 45$\pm$11\%, consistent with a sub-maximal disc. The best
fitting model also allows us to derive the global properties of the
galaxy. The virial mass of the dark matter halo (calculated in a radius with an average density of 200$\rho_{\rm crit}$) is
(2.0$\pm$0.6)$\times$10$^{12}$M$_\odot$ (concentration parameter, $c=r_{200}/r_h=13.3$), the stellar mass is
(7.4$\pm$0.3)$\times$10$^{10}$M$_\odot$ for the bulge and
(7.6$\pm$1.9)$\times$10$^{10}$M$_\odot$ for the disc. As expected for
this particular lensing configuration, the bulge component is very
well constrained, but interesting results can be obtained on the disc
as well. The stellar to total mass ratio is $7.0\pm2.3$ \% in good
agreement with typical numbers for galaxies of comparable mass.

\section{Discussion and comparison with previous work}\label{discussion}

The constraints used to model the galaxy in \citet{trott02} were image
positions and the two neutral hydrogen rotation points at large radius
(and well away from the image positions). Given that no kinematic
information was available in the central regions, the inferred
velocity curve due to the four mass components was assumed to be
circular. This means that an acceptable fit to the data was attainable
with the four smoothly modelled mass components because the
constraints effectively lie at only two radii --- 0.9$\arcsec$
(images) and $\sim$30$\arcsec$ (HI data). There were considerable
degeneracies in the model at other radii \citep[e.g. see Figure 2
from][]{trott04}.

In the present analysis, the addition of kinematic data provides
additional constraints out to a radius of $\sim$5$\arcsec$ ($\sim$4
kpc). As displayed in Figure \ref{best_fit_rc_no_prior}, constraining
information is available in the central 5$\arcsec$ and at two outer
points in the galaxy, leaving the models to fit the remainder of radii
according to their input density profiles. The bulge is
well-constrained because it contributes considerably to the mass
within the inner 5$\arcsec$. The mass-to-light ratios of this study
are more tightly constrained than those of \citet{trott02} --- their
1$\sigma$ uncertainties were calculated while holding the values of
other parameters at their minimum, thereby underestimating the true
uncertainties. The bulge mass-to-light ratio is consistent with
previous measurements from early-type galaxies
M/L$_B$=7.9$\pm$2.3$\Upsilon_{\odot,h=0.70}$. \citep{gerhard01,treu04}.
Similarly, \citet{jiang07} report M/L$_B$ =
7.2$\pm$0.5$\Upsilon_{\odot,h=0.70}$ from an analysis of 22 lens
galaxies.  As far as the disc mass-to-light ratio is concerned, our
best fitting value is also consistent with typical measurements
reported in the literature \citep[e.g.][and references
therein]{courteau99}.

The dark matter halo, however, contributes mostly in the outer regions
of the galaxy where few constraints are available. In addition, unlike
the bulge and disc, the absence of a light profile to guide the
density profile of the dark matter yields additional freedom in the
model. This allows a more flexible model, but also increases
degeneracy. It is thus remarkable to see how well the dark matter
halo is constrained by the available data.  However, more detailed
information about the structure of the dark matter halo would require
information at radii greater than 6$\arcsec$, where its influence
increases. Alternatively, to gain more information about the halo from
these data, one could place a prior on the bulge and disc
mass-to-light ratios based on studies of other galaxies (or on stellar
synthesis models) and determine whether such constraints narrowed the
allowed value of the inner slope. This is left for future work.

The best-fitting solution described in \citet{trott02} had a more massive disc and less massive halo compared with the best-fitting solution described here. The bulge mass was consistent between the two studies, within quoted errors. The errors quoted in \citet{trott02} were calculated assuming no parameter covariances, and therefore underestimate the true errors as calculated by marginalising over the remaining parameters (as performed in this work). Therefore it is difficult to determine whether the halo and disc masses (and therefore the disc mass-to-light ratio) are consistent between the two studies. Note, however, that the dark matter halo mass profile used in this study was different to that in \citet{trott02}, allowing freedom in the inner slope.

The fit to the image positions in this study is not as good as the fit in \citet{trott02}. This is due to the increased complexity of the model in this work, and the need for the fit to balance the image positions and the kinematics. In \citet{trott02}, the image positions formed the most precise portion of the data (the HI points, being far from the galactic centre and with relatively large error bars are easier to fit), and therefore were able to be well-fitted. In this work, the first and second order kinematics formed a relatively larger proportion of the $\chi^2$ calculation, and therefore were fitted at the expense of the image positions.

In future work the modelling could be improved to include any effects of external shear (G2237+0305 has other less massive systems identified along the line-of-sight) and to include the kinematic effects of the bar more consistently. The former could improve the lensing component of the $\chi^2$ without affecting the dynamical fit. In addition, the dark matter halo could be modelled to be non-spherical with flattening aligned with that of the disc (as suggested by studies of other lensing systems).

Despite the comparatively good model presented in this present work,
it is important to keep in mind the following caveats. The lens galaxy
2237+0305 has additional structure that is not accounted for in the
four component model, e.g. prominent spiral arms that are attached to
the bar component. In addition, the bar is not treated in a
self-consistent manner.  These may not be important for the lensing
but still produce anisotropy in the kinematics. The bar is difficult
to model kinematically and an attempt has not been made in this
work. However, in the central few arcseconds its presence is small and
we opted to neglect it for the inner kinematics, including its
effect only in the lensing and for the overall rotation curve.

The image magnifications and total magnification are dependent on the
balance between the steepness and the anisotropy of the overall mass
model. As demonstrated by \citet{wambsganss94}, 2237+0305 allows a
range of models with degeneracy between the density slope and
ellipticity. Inclusion of kinematics in our model has reduced that
degeneracy. The magnification ratios are consistent with the mid-IR
measurements of \citet{agol00}. They argued that mid-IR
measurements are the least likely to be contaminated by stellar
microlensing, electron scattering and extinction \citep[see
also][]{chiba05,agol06}.

In terms of global properties, the central galaxy is bulge-dominated,
and modelling the mass distribution of the bulge accurately is most
important for an accurate overall model. Beyond $r\sim$20$\arcsec$,
the halo appears to continue to increase in rotational support in the
outer regions, and to absorb the rotation beyond the optical radius.

Our results can be compared with those of \citet{vandenven08}. Their
measured central velocity dispersion ($166\pm2$ kms$^{-1}$) is
consistent with that presented in this work ($172\pm9$ kms$^{-1}$)
within the errors. Although they had the advantage of two-dimensional
data and the additional information that accompanies that, our data
are based on stellar kinematics from the Mgb-Fe region, observed in a
cleaner region of the sky than the CaII triplet targeted by their
observations. This allowed us to reach further out in projected
distance from the centre, albeit along preferred directions.  The
results of \citet{vandenven08} are complementary to ours: they fit a
fourier-based lens model, fitting the image positions and radio flux
ratios, and combine the information with the implied mass distribution
from the kinematics and light distribution, to obtain an estimate of
the luminous mass-to-light ratio. A comparison of this value with that
obtained using a single stellar population model yields a maximum dark
matter contribution within the Einstein radius of 20\%, which is higher than our best fit value of 7\% and our upper limit of 15\%
(3$\sigma$). In addition, their comparison of the lens-based mass
model and the observed luminosity distribution indicates that mass
follows light in the inner regions, and any halo contribution is
relatively constant. Our analysis focuses on determining a simple yet
comprehensive description of the galaxy over a broader range of
scales. Although the results for the dark matter halo can be improved
by more extended and precise kinematic data, we achieved our aim of
determining the bulge and disc mass-to-light ratios with good
precision, without the need to use stellar population synthesis
models. Future studies with the addition of stellar population
synthesis models will be very valuable.

The good agreement between our results and those of
\citet{vandenven08} -- based on largely independent datasets (except
for the strong lensing) and methodologies -- is encouraging: the
combination of lensing and dynamics is a powerful tool to study spiral
galaxies, not only elliptical galaxies. As larger samples of spiral
lens galaxies are being discovered by the SLACS \citep{bolton08a} and
other surveys -- including many where the Einstein radius will be
equal or larger than the scale length of the disc -- the combination
of lensing, kinematics, and stellar population synthesis models should
enable substantial progress in our understanding of the structure of
disc galaxies, and therefore their formation and evolution.

\section{Summary}\label{conclusion}

Data from the ESI echelle instrument on the Keck II telescope were
used to measure the major and minor axis kinematics of the spiral lens
galaxy 2237+0305. The rotation and velocity dispersion profile in the
central few arcseconds of 2237+0305 are combined with the lensed-image
positions and the rotation curve beyond the optical disc (HI rotation
points) to construct a structural and kinematic model of
2237+0305. Four components are used to constrain the mass distribution
--- a dark matter halo with variable inner logarithmic slope and scale
length; a de Vaucouleurs bulge, modelled structurally with a
flattening consistent with the light profile; an exponential disc,
inclined at 60$\degree$; a Ferrers bar, at a position angle
39$\degree$ from the disc major axis. The inner galaxy kinematics were
modelled with an axisymmetric two-integral model composed of the
bulge, disc, and dark matter halo.

The best-fitting solution shows that the lens galaxy is
bulge-dominated in the inner regions, with a massive stellar disc and
dark matter halo that provide rotational support beyond the bulge. The
bulge and disc B-band mass-to-light ratios
(M/L$_b$=6.6$\pm$0.3$\Upsilon_\odot$,
M/L$_d$=1.2$\pm$0.3$\Upsilon_\odot$) are consistent with typical
values found for local galaxies. The bulge mass is very well
constrained by the combination of lensing and dynamics. The disc mass
is less well constrained because of residual degeneracies with the
halo mass, although the best solutions indicate a sub-maximal disc,
contributing $45\pm11$ \% of the support at 2.2 scale lengths.  The
inner logarithmic slope of the dark matter halo is consistent with a
negative inner logarithmic slope of $\gamma=1$, with a preferred value
of $\gamma$=0.9$\pm$0.3 (note that this may still be dependent on our use of an axisymmetric two-integral kinematic model).

This work illustrates the potential of lensing and dynamics to
investigate the internal structure of spiral galaxies \citep[see
also][]{vandenven08}. The system Q2237+0305 was the natural starting
point for this kind of investigation due to the unusually low redshift
of the lens galaxy. It is encouraging that such a complex galaxy,
including a bar in the central regions, can be described with a
relatively simple model described here over a wide range of scales. A
partial limitation of this target, however, is that the multiple
images are located at a particularly small physical scale --
significantly smaller than the effective radius of the bulge and the
exponential scale length of the disc -- due once again to the
unusually low redshift of the lens galaxy. This implies that the
lensing geometry has relatively little leverage on the dark matter
halo, compared to typical lenses at higher redshifts where the
Einstein radius can be up to a few half-light radii, where the dark
matter halo starts to be a substantial fraction of the total mass
\citep[e.g.][]{treu04}. Current and future samples of disc lens
galaxies are likely to be found at higher redshift, thus providing an
opportunity to break the disc-halo degeneracy even further, as well as
to investigate evolutionary trends.

\section*{Acknowledgments}
We wish to thank the anonymous referee for improving the manuscript with suggestions that clarify our methodology.
CMT acknowledges the support provided by the David Hay Memorial Fund
during preparation of this manuscript.  L.V.E.K. is supported (in
part) through an NWO-VIDI program subsidy (project number
639.042.505).  T.T.  acknowledges support from the NSF through CAREER
award NSF-0642621, by the Sloan Foundation through a Sloan Research
Fellowship, and by the Packard Foundation through a Packard
Fellowship.  Support for HST archival program \#9960 was provided by
NASA through a grant from the Space Telescope Science Institute, which
is operated by the Association of Universities for Research in
Astronomy, Inc., under NASA contract NAS 5-26555. We thank Aaron
Dutton, Randall Wayth and Phil Marshall for many insightful conversations.

\appendix

\section{Data reduction}\label{data_reduction}

The ESI observations require some special preparation and calibration
before they can be combined and used for kinematic analysis. The
calibration process undertaken includes the following steps: bias
subtraction, flat-fielding, rectification, cosmic-ray rejection and
sky subtraction. These steps were performed by the package {\tt
EASI2D}, which is developed by David J.~Sand and T.~Treu
\citep{sand04} for easy extraction of echelle orders.

\begin{figure}
\begin{center}
\includegraphics[width=8cm]{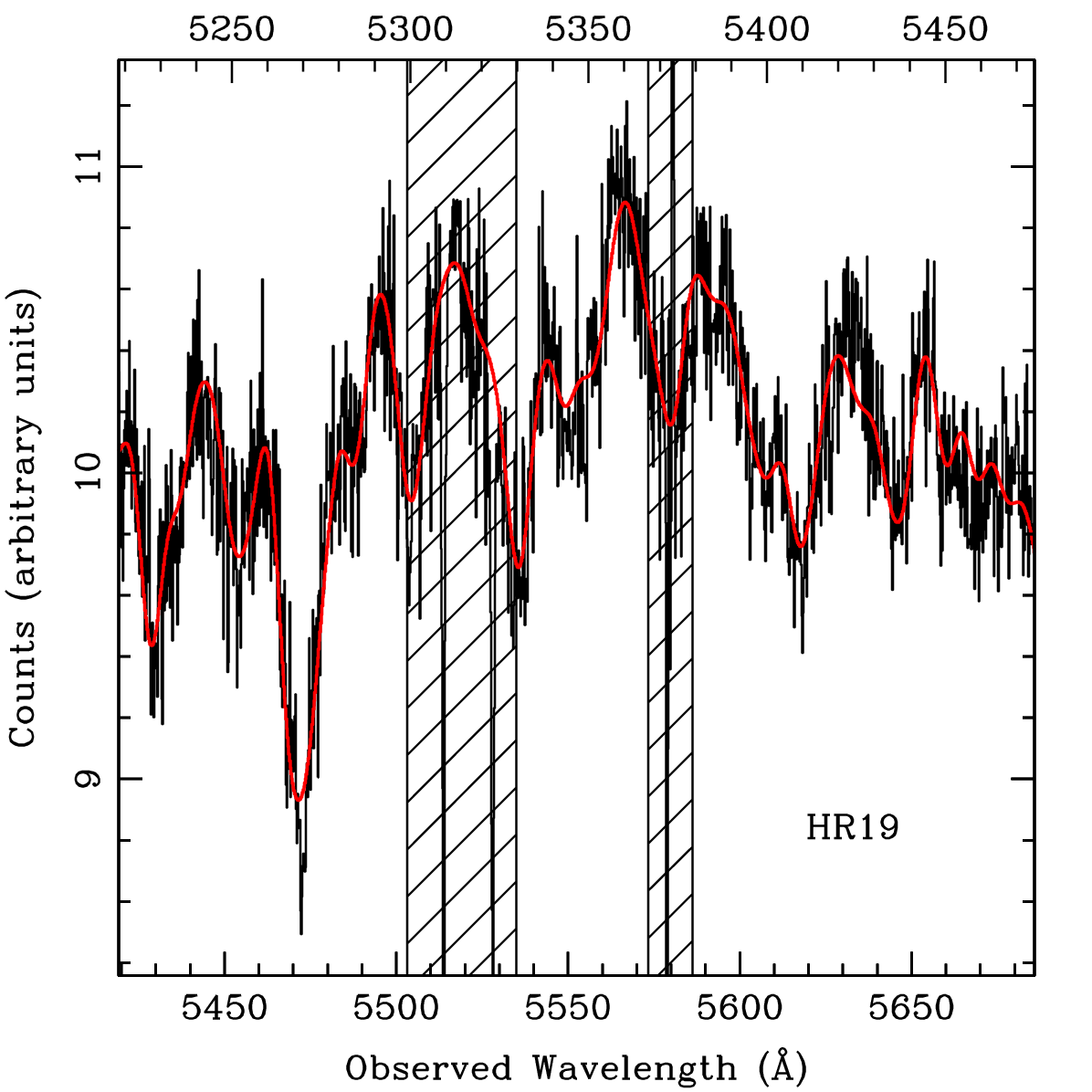}
\caption{Best kinematic fit in the spectral region of the Mg and Fe
absorption lines (rest wavelengths are indicated on the top axis of
the plot). The data are shown as a black histogram, while the best
fitting template is shown as a red curve. Spectral regions affected by
intervening absorption line systems are masked out during the fit and
shown as shaded areas.
\label{best}}
\end{center}
\end{figure}

Given the lack of emission lines in the spectra of the old stellar
population, strong absorption features need to be used in the
kinematic analysis. The canonical region for stellar absorption
kinematics in the optical -- the Mgb-Fe complex around 5200 \AA\, rest
frame shown in Figure~\ref{best} -- was used as our primary kinematic
dataset. As an additional test, we also fitted the NaD doublet at
5892\AA, which is generally considered less reliable because of
possible interstellar absorption~\citep{sparks97}.

In practice, the Gauss-Hermite Pixel Fitting Software developed by
R. P. van der Marel
(http://www-int.stsci.edu/$\sim$marel/software/pixfit.html) is
employed to determine the kinematic properties along both the major
and minor axes. The template and galaxy spectra are prepared to have
the same resolution, number of pixels and wavelength range. They are
then compared using the iterative fitting method described in
\citet{vandermarel94}. The variable parameters used to find the
best-fitting (lowest $\chi^2$) solution are the order of the
polynomial fit to the continuum and the spectral type used as the
template. The rotation curves were found to be quite robust to
parameter variations, while as expected the velocity dispersion
profile requires more care in choosing the parameters.  The region
shown in Figure~\ref{best} was found to be a good compromise between
using a large enough region and avoiding as much as possible the broad
AGN features. A large number of tests were run to explore spectral
ranges, masking areas and polynomial order for continuum fitting. Our
error bars include all these sources of errors which dominate over the
random errors.

The radius range is also extended beyond the region where the galaxy
signal is obvious in order to include as much information as can be
obtained from the data.  The data are binned radially to 0.6$\arcsec$
bins, in order for the points to be independent given the seeing. The
outer points do not produce good fits --- beyond $r{\sim}4\arcsec$,
the galaxy signal is lost in the noise.  The best-fitting stellar
templates are from Giant stars with type G9III -- K5III, consistent
with the imaging results of~\citet{yee88} and produce statistically
indistinguishable fits. Figure~\ref{best} displays an example fit. The
stability and accuracy of the kinematic results are very good --- all
templates produce mutually consistent fits and the residuals appear
consistent with noise.

The NaD region gave consistent rotation curves to the Mgb-Fe region,
but systematically higher velocity dispersion profiles by
25\%. Possible explanations of this difference -- in addition to the
effects of interstellar absorption -- are contamination by quasar
light and the diversity of possibly kinematically distinct stellar
populations in the targeted regions. Although we feel that our choice
of the Mgb-Fe region is justified, we repeated our analysis using the
NaD-based stellar velocity dispersion profile and found our results to
change only marginally. None of the conclusions of this paper is
affected significantly by this choice.

The centre of the galaxy is difficult to locate from the data due to
the significant contamination from the lensed quasars and the
difficulties associated with image deconvolution. We therefore
symmetrise the data to match the shape of the curve on both sides to
find the kinematic centre.  Once the centre is found, the rotation
curves are symmetric within errors. The kinematic profiles obtained
with the four best fitting templates have been averaged to produce the
final major axis results. The analysis of the minor axis data is
similar to that performed for the major axis.

The central line-of-sight velocity dispersion is consistent within
$1\sigma$ uncertainties between the major and minor axis: $\sigma_c
\simeq$ 172$\pm$9 kms$^{-1}$, which compares well with the result of
\citet{vandenven08} of $\sigma_c \simeq$ 166$\pm$2 kms$^{-1}$. Our
uncertainties are larger, due to our inclusion of dominant systematic
errors such as template and continuum mismatch. Our results are
marginally lower than the value inferred by \citet{foltz92}
($215\pm30$ kms$^{-1}$), possibly for the same reasons that cause the
mismatch between the Mgb-Fe and NaD results.

\section{Two-integral axisymmetric galaxy models}\label{axisymmetric}

Here we briefly summarise the method of \citet{vandermarel07} to
produce axisymmetric dynamical models from photometric data. See the
original paper for further details.

We assume the galaxy can be fitted with an axisymmetric model, and
that there are two integrals of motion. In this regime, the two
coupled Jean's equations are:
\begin{eqnarray}
\frac{\partial{\rho}\overline{v_z^2}}{\partial{z}} + \rho\frac{\partial\Phi}{\partial{z}} = 0, \\\label{jeanseqns1}
\frac{\partial{\rho}\overline{v_R^2}}{\partial{R}} + \rho\frac{\partial\Phi}{\partial{R}} + \frac{\rho}{R}\left[{\overline{v_R^2}-\overline{v_{\phi}^2}}\right] = 0,\label{jeanseqns2}
\end{eqnarray}
where $\overline{v_R^2}\equiv\overline{v_z^2}$.

The gradients of the gravitational potential, $\nabla\Phi$, are calculated from the currently used mass density profiles of the mass components, according to \citep{bintre87}:
\begin{equation}
{\bf{F}} = -\nabla\Phi = -\pi{G}\sum_{i=1}^N\sqrt{1-e_i^2}a_i{\int^\infty_0}\frac{\rho_i(m^2)\nabla{m^2}}{(\tau+a_i^2)\sqrt(\tau+b_i^2)}d\tau,
\end{equation}
where,
\begin{equation}
m^2/a_i^2 = \frac{R^2}{\tau+a_i^2} + \frac{z^2}{\tau+b_i^2},
\end{equation}
and $a_i, b_i$ are the semi-major and -minor axes of the mass component, and the summation is over all distinct mass components.

The mean streaming motion is not constrained by the Jean's equations, and \citet{vandermarel07} introduce a constrained scaling parameter, $k$, to control the amount of rotation,
\begin{equation}\label{streaming}
\overline{v_\phi} = k\sqrt{\overline{v^2_\phi} - \overline{v^2_R}}.
\end{equation}
The value of $k$ is constrained by the necessity of positive dispersion everywhere,
\begin{equation}
k \leq \min_{(R,z)} [\overline{v^2_\phi}/(\overline{v^2_\phi}-\overline{v^2_R})]^{1/2}.
\end{equation}

The mass components and velocity moments then need to be inclined to the viewing angle, and the projected line-of-sight moments determined according to,
\begin{equation}
\langle\overline{v_w^m}\rangle(x,y) = \frac{1}{\Sigma(x,y)}\int^{\infty}_{-\infty}\rho(x,y,w)\overline{v_w^m}(x,y,w)dw,
\end{equation}
where $w$ denotes the line-of-sight direction, and,
\begin{eqnarray}
\overline{v_w} &=& \overline{v_\phi}\cos{\phi}\sin{i},\\\nonumber
\overline{v_w^2} &=& (\overline{v_\phi^2} - \overline{v_R^2})\cos^2\phi\sin^2i + \overline{v_R^2}
\end{eqnarray}
are the first and second moments, and $i$ is the inclination angle. The rotation component, $V$, and velocity dispersion, $\sigma$, are then derived as,
\begin{eqnarray}
V &=& <\overline{v_w}>,\\\nonumber
\sigma^2 &=& V^2 - \sigma^2_{RMS}\\
&=& V^2 - <\overline{v_w^2}>.
\end{eqnarray}

The prescription to determine the first and second velocity moments using this method is as follows:
\begin{itemize}
\item For a given mass model, calculate the gravitational potential derivatives and density profiles on a ($R$, $z$) grid;
\item Calculate $\overline{v_z^2} = \overline{v^2_R}$ on the same grid according to \ref{jeanseqns1};
\item Calculate $\overline{v^2_\phi}$ on the same grid according to \ref{jeanseqns2};
\item Calculate the mean streaming motion on the same grid, according to \ref{streaming}, and ensure it meets the positivity requirement;
\item Incline the density distributions of the mass components to the line-of-sight;
\item Calculate the line-of-sight first and second velocity moments;
\item Convolve the moments with seeing, slit width and pixel size.
\end{itemize}

\bibliographystyle{mn2e}
\bibliography{trott_cm_astroph}

\begin{thebibliography}{}

\bibitem[\protect\citeauthoryear{{Agol}, {Jones} \& {Blaes}}{{Agol}
  et~al.}{2000}]{agol00}
{Agol} E.,  {Jones} B.,    {Blaes} O.,  2000, ApJ, 545, 657

\bibitem[\protect\citeauthoryear{{Agol} \& {Kochanek}}{{Agol} \&
  {Kochanek}}{2006}]{agol06}
{Agol} E.,  {Kochanek} C.,  2006, in Bulletin of the American Astronomical
  Society Vol.~38 of Bulletin of the American Astronomical Society, {Mid
  Infrared Observations of Quasar Lenses}.
pp 928--+

\bibitem[\protect\citeauthoryear{{Bertin}, {Ciotti} \& {Del Principe}}{{Bertin}
  et~al.}{2002}]{bertin02}
{Bertin} G.,  {Ciotti} L.,    {Del Principe} M.,  2002, A\&A, 386, 149

\bibitem[\protect\citeauthoryear{{Binney} \& {Tremaine}}{{Binney} \&
  {Tremaine}}{1987}]{bintre87}
{Binney} J.,  {Tremaine} S.,  1987, {Galactic Dynamics}.
Princeton, NJ, Princeton University Press

\bibitem[\protect\citeauthoryear{{Bolton}, {Burles}, {Koopmans}, {Treu},
  {Gavazzi}, {Moustakas}, {Wayth} \& {Schlegel}}{{Bolton}
  et~al.}{2008}]{bolton08a}
{Bolton} A.~S.,  {Burles} S.,  {Koopmans} L.~V.~E.,  {Treu} T.,  {Gavazzi} R.,
  {Moustakas} L.~A.,  {Wayth} R.,    {Schlegel} D.~J.,  2008, ApJ, 682, 964

\bibitem[\protect\citeauthoryear{{Bolton}, {Treu}, {Koopmans}, {Gavazzi},
  {Moustakas}, {Burles}, {Schlegel} \& {Wayth}}{{Bolton}
  et~al.}{2008}]{bolton08b}
{Bolton} A.~S.,  {Treu} T.,  {Koopmans} L.~V.~E.,  {Gavazzi} R.,  {Moustakas}
  L.~A.,  {Burles} S.,  {Schlegel} D.~J.,    {Wayth} R.,  2008, ApJ, 684, 248

\bibitem[\protect\citeauthoryear{{Borriello} \& {Salucci}}{{Borriello} \&
  {Salucci}}{2001}]{borriello01}
{Borriello} A.,  {Salucci} P.,  2001, MNRAS, 323, 285

\bibitem[\protect\citeauthoryear{{Brada{\v c}}, {Clowe}, {Gonzalez},
  {Marshall}, {Forman}, {Jones}, {Markevitch}, {Randall}, {Schrabback} \&
  {Zaritsky}}{{Brada{\v c}} et~al.}{2006}]{bradac06}
{Brada{\v c}} M.,  {Clowe} D.,  {Gonzalez} A.~H.,  {Marshall} P.,  {Forman} W.,
   {Jones} C.,  {Markevitch} M.,  {Randall} S.,  {Schrabback} T.,    {Zaritsky}
  D.,  2006, ApJ, 652, 937

\bibitem[\protect\citeauthoryear{{Bucholz}}{{Bucholz}}{1969}]{bucholz69}
{Bucholz} H.,  1969, {The confluent hypergeometric function : with special
  emphasis on its applications }.
Springer-Verlag

\bibitem[\protect\citeauthoryear{{Chiba}, {Minezaki}, {Kashikawa}, {Kataza} \&
  {Inoue}}{{Chiba} et~al.}{2005}]{chiba05}
{Chiba} M.,  {Minezaki} T.,  {Kashikawa} N.,  {Kataza} H.,    {Inoue} K.~T.,
  2005, ApJ, 627, 53

\bibitem[\protect\citeauthoryear{{Courteau} \& {Rix}}{{Courteau} \&
  {Rix}}{1999}]{courteau99}
{Courteau} S.,  {Rix} H.,  1999, ApJ, 513, 561

\bibitem[\protect\citeauthoryear{{Crane}, {Albrecht}, {Barbieri} \& {et
  al.}}{{Crane} et~al.}{1991}]{crane91}
{Crane} P.,  {Albrecht} R.,  {Barbieri} C.,    {et al.} 1991, ApJL, 369, L59

\bibitem[\protect\citeauthoryear{{Czoske}, {Barnab{\`e}}, {Koopmans}, {Treu} \&
  {Bolton}}{{Czoske} et~al.}{2008}]{czoske08}
{Czoske} O.,  {Barnab{\`e}} M.,  {Koopmans} L.~V.~E.,  {Treu} T.,    {Bolton}
  A.~S.,  2008, MNRAS, 384, 987

\bibitem[\protect\citeauthoryear{{de Blok}}{{de Blok}}{2005}]{deblok05}
{de Blok} W.~J.~G.,  2005, ApJ, 634, 227

\bibitem[\protect\citeauthoryear{{de Blok}, {McGaugh} \& {Rubin}}{{de Blok}
  et~al.}{2001}]{deblok01}
{de Blok} W.~J.~G.,  {McGaugh} S.~S.,    {Rubin} V.~C.,  2001, AJ, 122, 2396

\bibitem[\protect\citeauthoryear{{Dutton}, {Courteau}, {de Jong} \&
  {Carignan}}{{Dutton} et~al.}{2005}]{dutton05}
{Dutton} A.~A.,  {Courteau} S.,  {de Jong} R.,    {Carignan} C.,  2005, ApJ,
  619, 218

\bibitem[\protect\citeauthoryear{{Dutton}, {van den Bosch}, {Dekel} \&
  {Courteau}}{{Dutton} et~al.}{2007}]{dutton07}
{Dutton} A.~A.,  {van den Bosch} F.~C.,  {Dekel} A.,    {Courteau} S.,  2007,
  ApJ, 654, 27

\bibitem[\protect\citeauthoryear{{Eigenbrod}, {Courbin}, {Sluse}, {Meylan} \&
  {Agol}}{{Eigenbrod} et~al.}{2008a}]{eigenbrod08b}
{Eigenbrod} A.,  {Courbin} F.,  {Sluse} D.,  {Meylan} G.,    {Agol} E.,  2008a,
  in Proceedings of the Manchester Microlensing Conference: The 12th
  International Conference and ANGLES Microlensing Workshop. {``Microlensing
  variability in the gravitationally lensed quasar Q2237+0305 = Einstein
  Cross"}

\bibitem[\protect\citeauthoryear{{Eigenbrod}, {Courbin}, {Sluse}, {Meylan} \&
  {Agol}}{{Eigenbrod} et~al.}{2008b}]{eigenbrod08a}
{Eigenbrod} A.,  {Courbin} F.,  {Sluse} D.,  {Meylan} G.,    {Agol} E.,  2008b,
  A\&A, 480, 647

\bibitem[\protect\citeauthoryear{{Foltz}, {Hewett}, {Webster} \&
  {Lewis}}{{Foltz} et~al.}{1992}]{foltz92}
{Foltz} C.~B.,  {Hewett} P.~C.,  {Webster} R.~L.,    {Lewis} G.~F.,  1992, 386,
  L43

\bibitem[\protect\citeauthoryear{{Gerhard}, {Kronawitter}, {Saglia} \&
  {Bender}}{{Gerhard} et~al.}{2001}]{gerhard01}
{Gerhard} O.,  {Kronawitter} A.,  {Saglia} R.~P.,    {Bender} R.,  2001, AJ,
  121, 1936

\bibitem[\protect\citeauthoryear{{Gil-Merino} \& {Lewis}}{{Gil-Merino} \&
  {Lewis}}{2005}]{gilmerino05}
{Gil-Merino} R.,  {Lewis} G.~F.,  2005, A\&A, 437, L15

\bibitem[\protect\citeauthoryear{{Gorenstein}, {Shapiro} \&
  {Falco}}{{Gorenstein} et~al.}{1988}]{gorenstein88}
{Gorenstein} M.~V.,  {Shapiro} I.~I.,    {Falco} E.~E.,  1988, ApJ, 327, 693

\bibitem[\protect\citeauthoryear{{Grillo}, {Lombardi}, {Rosati}, {Bertin},
  {Gobat}, {Demarco}, {Lidman}, {Motta} \& {Nonino}}{{Grillo}
  et~al.}{2008}]{grillo08}
{Grillo} C.,  {Lombardi} M.,  {Rosati} P.,  {Bertin} G.,  {Gobat} R.,
  {Demarco} R.,  {Lidman} C.,  {Motta} V.,    {Nonino} M.,  2008, A\&A, 486, 45

\bibitem[\protect\citeauthoryear{{Halkola}, {Seitz} \& {Pannella}}{{Halkola}
  et~al.}{2006}]{halkola06}
{Halkola} A.,  {Seitz} S.,    {Pannella} M.,  2006, MNRAS, 372, 1425

\bibitem[\protect\citeauthoryear{{Hasan} \& {Burrows}}{{Hasan} \&
  {Burrows}}{1995}]{hasan95}
{Hasan} H.,  {Burrows} C.~J.,  1995, PASP, 107, 289

\bibitem[\protect\citeauthoryear{{Hayashi}, {Navarro}, {Power}, {Jenkins},
  {Frenk}, {White}, {Springel}, {Stadel} \& {Quinn}}{{Hayashi}
  et~al.}{2004}]{hayashi04}
{Hayashi} E.,  {Navarro} J.~F.,  {Power} C.,  {Jenkins} A.,  {Frenk} C.~S.,
  {White} S.~D.~M.,  {Springel} V.,  {Stadel} J.,    {Quinn} T.~R.,  2004,
  MNRAS, 355, 794

\bibitem[\protect\citeauthoryear{{Huchra}, {Gorenstein}, {Kent}, {Shapiro},
  {Smith}, {Horine} \& {Perley}}{{Huchra} et~al.}{1985}]{huchra85}
{Huchra} J.,  {Gorenstein} M.,  {Kent} S.,  {Shapiro} I.,  {Smith} G.,
  {Horine} E.,    {Perley} R.,  1985, AJ, 90, 691

\bibitem[\protect\citeauthoryear{{Jiang} \& {Kochanek}}{{Jiang} \&
  {Kochanek}}{2007}]{jiang07}
{Jiang} G.,  {Kochanek} C.~S.,  2007, ApJ, 671, 1568

\bibitem[\protect\citeauthoryear{{Keeton}}{{Keeton}}{2001}]{keeton01}
{Keeton} C.~R.,  2001, astro-ph/0102341

\bibitem[\protect\citeauthoryear{{Kent} \& {Falco}}{{Kent} \&
  {Falco}}{1988}]{kent88}
{Kent} S.~M.,  {Falco} E.~E.,  1988, AJ, 96, 1570

\bibitem[\protect\citeauthoryear{{Kochanek}}{{Kochanek}}{1991}]{kochanek91}
{Kochanek} C.~S.,  1991, ApJ, 373, 354

\bibitem[\protect\citeauthoryear{{Koopmans}, {de Bruyn} \&
  {Jackson}}{{Koopmans} et~al.}{1998}]{koopmans98}
{Koopmans} L.~V.~E.,  {de Bruyn} A.~G.,    {Jackson} N.,  1998, MNRAS, 295, 534

\bibitem[\protect\citeauthoryear{{Koopmans} \& {Treu}}{{Koopmans} \&
  {Treu}}{2002}]{koopmans02}
{Koopmans} L.~V.~E.,  {Treu} T.,  2002, 568, L5

\bibitem[\protect\citeauthoryear{{Koopmans} \& {Treu}}{{Koopmans} \&
  {Treu}}{2003}]{koopmans03}
{Koopmans} L.~V.~E.,  {Treu} T.,  2003, ApJ, 583, 606

\bibitem[\protect\citeauthoryear{{Koopmans}, {Treu}, {Bolton}, {Burles} \&
  {Moustakas}}{{Koopmans} et~al.}{2006}]{koopmans06}
{Koopmans} L.~V.~E.,  {Treu} T.,  {Bolton} A.~S.,  {Burles} S.,    {Moustakas}
  L.~A.,  2006, ApJ, 649, 599

\bibitem[\protect\citeauthoryear{{Lewis}, {Irwin}, {Hewett} \& {Foltz}}{{Lewis}
  et~al.}{1998}]{lewis98}
{Lewis} G.~F.,  {Irwin} M.~J.,  {Hewett} P.~C.,    {Foltz} C.~B.,  1998, MNRAS,
  295, 573

\bibitem[\protect\citeauthoryear{{Lima Neto}, {Gerbal} \& {M{\' a}rquez}}{{Lima
  Neto} et~al.}{1999}]{limaneto99}
{Lima Neto} G.~B.,  {Gerbal} D.,    {M{\' a}rquez} I.,  1999, MNRAS, 309, 481

\bibitem[\protect\citeauthoryear{{Limousin}, {Richard}, {Jullo}, {Kneib},
  {Fort}, {Soucail}, {El{\'{\i}}asd{\'o}ttir}, {Natarajan}, {Ellis}, {Smail},
  {Czoske}, {Smith}, {Hudelot}, {Bardeau}, {Ebeling}, {Egami} \&
  {Knudsen}}{{Limousin} et~al.}{2007}]{limousin07}
{Limousin} M.,  {Richard} J.,  {Jullo} E.,  {Kneib} J.-P.,  {Fort} B.,
  {Soucail} G.,  {El{\'{\i}}asd{\'o}ttir} {\'A}.,  {Natarajan} P.,  {Ellis}
  R.~S.,  {Smail} I.,  {Czoske} O.,  {Smith} G.~P.,  {Hudelot} P.,  {Bardeau}
  S.,  {Ebeling} H.,  {Egami} E.,    {Knudsen} K.~K.,  2007, ApJ, 668, 643

\bibitem[\protect\citeauthoryear{{{\L}okas}, {Mamon} \& {Prada}}{{{\L}okas}
  et~al.}{2005}]{lokas04}
{{\L}okas} E.~L.,  {Mamon} G.~A.,    {Prada} F.,  2005, MNRAS, 363, 918

\bibitem[\protect\citeauthoryear{{Maller}, {Simard}, {Guhathakurta}, {Hjorth},
  {Jaunsen}, {Flores} \& {Primack}}{{Maller} et~al.}{2000}]{maller00}
{Maller} A.~H.,  {Simard} L.,  {Guhathakurta} P.,  {Hjorth} J.,  {Jaunsen}
  A.~O.,  {Flores} R.~A.,    {Primack} J.~R.,  2000, ApJ, 533, 194

\bibitem[\protect\citeauthoryear{{Mihov}}{{Mihov}}{2001}]{mihov01}
{Mihov} B.~M.,  2001, A\&A, 370, 43

\bibitem[\protect\citeauthoryear{{M{\"o}ller}, {Hewett} \&
  {Blain}}{{M{\"o}ller} et~al.}{2003}]{moller03}
{M{\"o}ller} O.,  {Hewett} P.,    {Blain} A.~W.,  2003, MNRAS, 345, 1

\bibitem[\protect\citeauthoryear{{Moore}, {Quinn}, {Governato}, {Stadel} \&
  {Lake}}{{Moore} et~al.}{1999}]{moore99}
{Moore} B.,  {Quinn} T.,  {Governato} F.,  {Stadel} J.,    {Lake} G.,  1999,
  MNRAS, 310, 1147

\bibitem[\protect\citeauthoryear{{Mu{\~ n}oz}, {Kochanek} \& {Keeton}}{{Mu{\~
  n}oz} et~al.}{2001}]{munoz01}
{Mu{\~ n}oz} J.~A.,  {Kochanek} C.~S.,    {Keeton} C.~R.,  2001, ApJ, 558, 657

\bibitem[\protect\citeauthoryear{{Navarro}, {Frenk} \& {White}}{{Navarro}
  et~al.}{1996}]{nfw96}
{Navarro} J.~F.,  {Frenk} C.~S.,    {White} S.~D.~M.,  1996, ApJ, 462

\bibitem[\protect\citeauthoryear{{Navarro}, {Hayashi}, {Power}, {Jenkins},
  {Frenk}, {White}, {Springel}, {Stadel} \& {Quinn}}{{Navarro}
  et~al.}{2004}]{nav04}
{Navarro} J.~F.,  {Hayashi} E.,  {Power} C.,  {Jenkins} A.~R.,  {Frenk} C.~S.,
  {White} S.~D.~M.,  {Springel} V.,  {Stadel} J.,    {Quinn} T.~R.,  2004,
  MNRAS, 349, 1039

\bibitem[\protect\citeauthoryear{{Noordermeer}}{{Noordermeer}}{2008}]{noorderm%
eer08}
{Noordermeer} E.,  2008, MNRAS, 385, 1359

\bibitem[\protect\citeauthoryear{{Peng}, {Ho}, {Impey} \& {Rix}}{{Peng}
  et~al.}{2002}]{peng02}
{Peng} C.~Y.,  {Ho} L.~C.,  {Impey} C.~D.,    {Rix} H.-W.,  2002, AJ, 124, 266

\bibitem[\protect\citeauthoryear{{P{\'e}rez}, {Fux} \& {Freeman}}{{P{\'e}rez}
  et~al.}{2004}]{perez04}
{P{\'e}rez} I.,  {Fux} R.,    {Freeman} K.,  2004, A\&A, 424, 799

\bibitem[\protect\citeauthoryear{{Rauch}, {Sargent}, {Barlow} \&
  {Simcoe}}{{Rauch} et~al.}{2002}]{rauch02}
{Rauch} M.,  {Sargent} W.~L.~W.,  {Barlow} T.~A.,    {Simcoe} R.~A.,  2002,
  ApJ, 576, 45

\bibitem[\protect\citeauthoryear{{Sackett}}{{Sackett}}{1997}]{sackett97}
{Sackett} P.~D.,  1997, ApJ, 483, 103

\bibitem[\protect\citeauthoryear{{Sand}, {Treu}, {Smith} \& {Ellis}}{{Sand}
  et~al.}{2004}]{sand04}
{Sand} D.~J.,  {Treu} T.,  {Smith} G.~P.,    {Ellis} R.~S.,  2004, ApJ, 604, 88

\bibitem[\protect\citeauthoryear{{Schmidt}}{{Schmidt}}{1996}]{schmidt96}
{Schmidt} R.~W.,  1996, Masters Thesis, pp~2--+

\bibitem[\protect\citeauthoryear{{Schneider}, {Turner}, {Gunn}, {Hewitt},
  {Schmidt} \& {Lawrence}}{{Schneider} et~al.}{1988}]{schneider88}
{Schneider} D.~P.,  {Turner} E.~L.,  {Gunn} J.~E.,  {Hewitt} J.~N.,  {Schmidt}
  M.,    {Lawrence} C.~R.,  1988, AJ, 96, 1755

\bibitem[\protect\citeauthoryear{{S\'{e}rsic}}{{S\'{e}rsic}}{1968}]{sersic68}
{S\'{e}rsic} J.~L.,  1968, {Atlas de galaxias australes}.
Cordoba, Argentina: Observatorio Astronomico

\bibitem[\protect\citeauthoryear{{Shin} \& {Evans}}{{Shin} \&
  {Evans}}{2007}]{shin07}
{Shin} E.~M.,  {Evans} N.~W.,  2007, MNRAS, 374, 1427

\bibitem[\protect\citeauthoryear{{Sparks}, {Carollo} \& {Macchetto}}{{Sparks}
  et~al.}{1997}]{sparks97}
{Sparks} W.~B.,  {Carollo} C.~M.,    {Macchetto} F.,  1997, ApJ, 486, 253

\bibitem[\protect\citeauthoryear{{Treu} \& {Koopmans}}{{Treu} \&
  {Koopmans}}{2004}]{treu04}
{Treu} T.,  {Koopmans} L.~V.~E.,  2004, ApJ, 611, 739

\bibitem[\protect\citeauthoryear{{Trott} \& {Webster}}{{Trott} \&
  {Webster}}{2002}]{trott02}
{Trott} C.~M.,  {Webster} R.~L.,  2002, MNRAS, 334, 621

\bibitem[\protect\citeauthoryear{{Trott} \& {Webster}}{{Trott} \&
  {Webster}}{2004}]{trott04}
{Trott} C.~M.,  {Webster} R.~L.,  2004, in IAU Symposium 220 {Determining the
  Properties of Galaxy 2237+0305 using Gravitational Lensing}.
pp 109--+

\bibitem[\protect\citeauthoryear{{Udalski}, {Szymanski}, {Kubiak},
  {Pietrzynski}, {Soszynski}, {Zebrun}, {Szewczyk}, {Wyrzykowski}, {Ulaczyk} \&
  {Wi{\^e}ckowski}}{{Udalski} et~al.}{2006}]{udalski06}
{Udalski} A.,  {Szymanski} M.~K.,  {Kubiak} M.,  {Pietrzynski} G.,  {Soszynski}
  I.,  {Zebrun} K.,  {Szewczyk} O.,  {Wyrzykowski} L.,  {Ulaczyk} K.,
  {Wi{\^e}ckowski} T.,  2006, Acta Astronomica, 56, 293

\bibitem[\protect\citeauthoryear{{Vakulik}, {Schild}, {Smirnov}, {Dudinov} \&
  {Tsvetkova}}{{Vakulik} et~al.}{2007}]{vakulik07}
{Vakulik} V.~G.,  {Schild} R.~E.,  {Smirnov} G.~V.,  {Dudinov} V.~N.,
  {Tsvetkova} V.~S.,  2007, MNRAS, 382, 819

\bibitem[\protect\citeauthoryear{{van Albada} \& {Sancisi}}{{van Albada} \&
  {Sancisi}}{1986}]{vanalbada86}
{van Albada} T.~S.,  {Sancisi} R.,  1986, Royal Society of London Philosophical
  Transactions Series A, 320, 447

\bibitem[\protect\citeauthoryear{{van de Ven}, {Falcon-Barroso}, {McDermid},
  {Cappellari}, {Miller} \& {de Zeeuw}}{{van de Ven}
  et~al.}{2008}]{vandenven08}
{van de Ven} G.,  {Falcon-Barroso} J.,  {McDermid} R.~M.,  {Cappellari} M.,
  {Miller} B.~W.,    {de Zeeuw} P.~T.,  2008, astro-ph/0807.4175

\bibitem[\protect\citeauthoryear{{van der Marel}}{{van der
  Marel}}{1994}]{vandermarel94}
{van der Marel} R.~P.,  1994, MNRAS, 270, 271

\bibitem[\protect\citeauthoryear{{van der Marel} \& {van Dokkum}}{{van der
  Marel} \& {van Dokkum}}{2007}]{vandermarel07}
{van der Marel} R.~P.,  {van Dokkum} P.~G.,  2007, ApJ, 668, 738

\bibitem[\protect\citeauthoryear{{Verheijen}, {Bershady}, {Swaters}, {Andersen}
  \& {Westfall}}{{Verheijen} et~al.}{2007}]{verheijen07}
{Verheijen} M.~A.~W.,  {Bershady} M.~A.,  {Swaters} R.~A.,  {Andersen} D.~R.,
   {Westfall} K.~B.,  2007, in Island Universes - Structure and Evolution of
  Disk Galaxies Astrophysics and Space Science Proceedings, {The Disk Mass
  Project: breaking the disk-halo degeneracy}.
pp 95--100

\bibitem[\protect\citeauthoryear{{Wambsganss} \& {Paczynski}}{{Wambsganss} \&
  {Paczynski}}{1994}]{wambsganss94}
{Wambsganss} J.,  {Paczynski} B.,  1994, AJ, 108, 1156

\bibitem[\protect\citeauthoryear{{Wayth}, {O'Dowd} \& {Webster}}{{Wayth}
  et~al.}{2005}]{wayth05}
{Wayth} R.~B.,  {O'Dowd} M.,    {Webster} R.~L.,  2005, MNRAS, 359, 561

\bibitem[\protect\citeauthoryear{{Windhorst}, {Burstein}, {Mathis},
  {Neuschaefer}, {Bertola}, {Buson}, {Koo}, {Matthews}, {Barthel} \&
  {Chambers}}{{Windhorst} et~al.}{1991}]{windhorst91}
{Windhorst} R.~A.,  {Burstein} D.,  {Mathis} D.~F.,  {Neuschaefer} L.~W.,
  {Bertola} F.,  {Buson} L.~M.,  {Koo} D.~C.,  {Matthews} K.,  {Barthel} P.~D.,
     {Chambers} K.~C.,  1991, ApJ, 380, 362

\bibitem[\protect\citeauthoryear{{Winn}, {Hall} \& {Schechter}}{{Winn}
  et~al.}{2003}]{winn03}
{Winn} J.~N.,  {Hall} P.~B.,    {Schechter} P.~L.,  2003, ApJ, 597, 672

\bibitem[\protect\citeauthoryear{{Wo{\'z}niak}, {Alard}, {Udalski},
  {Szyma{\'n}ski}, {Kubiak}, {Pietrzy{\'n}ski} \& {Zebru{\'n}}}{{Wo{\'z}niak}
  et~al.}{2000}]{wozniak00}
{Wo{\'z}niak} P.~R.,  {Alard} C.,  {Udalski} A.,  {Szyma{\'n}ski} M.,  {Kubiak}
  M.,  {Pietrzy{\'n}ski} G.,    {Zebru{\'n}} K.,  2000, ApJ, 529, 88

\bibitem[\protect\citeauthoryear{{Wucknitz}}{{Wucknitz}}{2002}]{wucknitz02}
{Wucknitz} O.,  2002, MNRAS, 332, 951

\bibitem[\protect\citeauthoryear{{Yee}}{{Yee}}{1988}]{yee88}
{Yee} H.~K.~C.,  1988, AJ, 95, 1331

\bibitem[\protect\citeauthoryear{{Yoo}, {Kochanek}, {Falco} \& {McLeod}}{{Yoo}
  et~al.}{2006}]{yoo06}
{Yoo} J.,  {Kochanek} C.~S.,  {Falco} E.~E.,    {McLeod} B.~A.,  2006, ApJ,
  642, 22

\bibitem[\protect\citeauthoryear{{Zackrisson}, {Bergvall}, {Marquart} \&
  {{\"O}stlin}}{{Zackrisson} et~al.}{2006}]{zackrisson06}
{Zackrisson} E.,  {Bergvall} N.,  {Marquart} T.,    {{\"O}stlin} G.,  2006,
  A\&A, 452, 857

\bibitem[\protect\citeauthoryear{{Zhao}}{{Zhao}}{1996}]{zhao96}
{Zhao} H.,  1996, MNRAS, 278, 488

\end{thebibliography}

\end{document}